# 1-D modelling and CFD study of the transient behaviour of a single phase Coupled Natural Circulation Loop

Akhil Dass D, Sateesh Gedupudi [1]

*Heat Transfer and Thermal Power Lab, Department of Mechanical Engineering, IIT Madras, Chennai 600036, India*

**Abstract**

A Coupled Natural Circulation Loop (CNCL) consists of two Natural Circulation Loops (NCL) coupled thermally via a common heat exchanger. The transient modelling of such systems that have practical relevance has not been reported in the literature to the best of the author's knowledge. The present work aims to bridge this gap and investigate the dynamic characteristics of a CNCL system using a 1-D mathematical model. The validation of the model is accomplished by comparison of the results obtained via 3-D CFD simulation. Both horizontal and vertical CNCL systems have been considered for this study and behaviour of the system for parallel and counter flow configurations in the heat exchanger section is elaborated. Transient and steady state CFD analysis has been conducted to analyse CNCL system for different heater and cooler orientations and flow initialisation. The behaviour of the CNCL system is then examined by carrying a thorough parametric study employing the validated 1-D single phase CNCL model with liquid sodium as the operating fluid. The CNCL orientation (vertical or horizontal) coupled with the heater and cooler configuration determines the system dynamics and behaviour. The CNCL system also exhibits chaotic flow oscillations at high heat loads.

## 1. Introduction

Natural circulation is a buoyancy-driven phenomenon that occurs in a closed conduit present in a body force field (gravitational, magnetic or centrifugal) when subjected to an external thermal stimulus. A Natural Circulation Loop (NCL) is a device with circular or rectangular geometry that operates based on this phenomenon when thermally stimulated at the heating and cooling sections (heat flux or internal heat generation or temperature boundary condition) of the loop, with the rest of the loop generally being insulated from the surroundings.

The past five decades have witnessed a surge in the amount of literature pertaining to NCLs, which can directly be attributed to their applications in several domains such as solar heaters, turbine blade cooling, geothermal energy extraction, nuclear power generation, electronic chip cooling, chemical process industries, closed loop pulsating heat pipes, refrigeration, ship propulsion etc. [1] .

Some general observations and characteristics of single-phase Natural Circulation Loops are listed below:

1. The NCL can be of any irregular geometry, the toroidal and rectangular geometries have been extensively studied because of their simplicity and practical relevance [1].
2. The heater and cooler orientations influence the transient and steady state behaviour of the NCL [2].
3. The NCL is a dynamical system, which exhibits chaotic behaviour at high power loads [3].
4. The NCL system is also very sensitive to the initial conditions at high power loads [3].

---

[1]Corresponding author. Tel.: +91 44 2257 4721, Email: sateeshg@iitm.ac.in

5. The power supplied to the NCL determines its dynamic characteristics. For lower power inputs, the NCL reaches a steady state, while increasing the power input leads to flow pulsations and/or reversals [3].

6. The NCL has a Lorenz like attractor for higher power inputs [3].

7. The viscous and buoyancy forces dictate the dynamic behaviour of the natural circulation system [4].

The above-mentioned characteristics and applications in numerous domains have augmented the research conducted in the field of NCLs. Despite the volume of research being conducted in the domain of natural circulation, very little attention has been paid to the subject of coupled/conjugate NCLs.

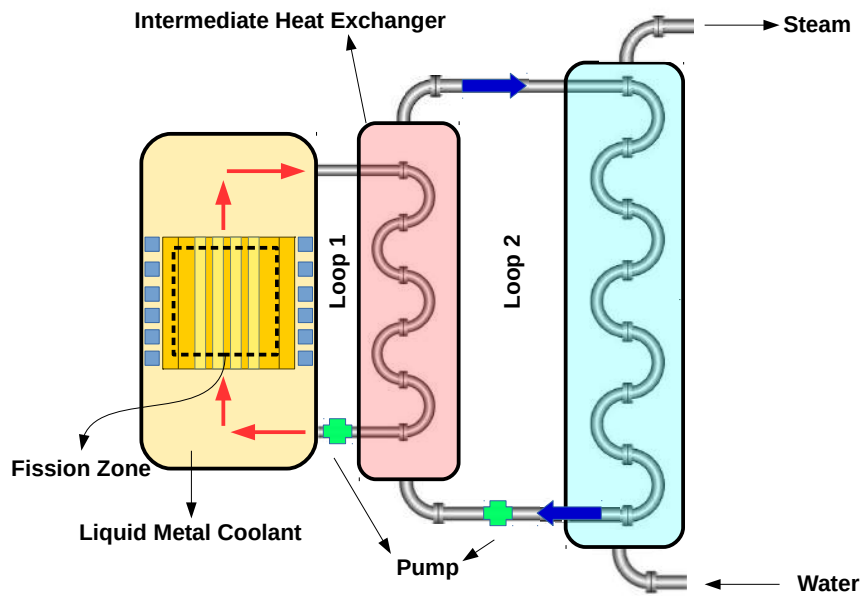

Figure 1: Schematic of the loop type LMFBR.

A Coupled Natural Circulation Loop (CNCL) is a device that is constructed from two NCLs, which are coupled thermally via a common heat exchanger section. The NCLs are not hydraulically linked; hence, the energy transfer between the two NCLs is solely due to thermal coupling at the heat exchanger section. Figure 1 represents a schematic of an LMFBR system, which acts as a CNCL when the pumps fail or they are switched off. Davis and Roppo [5] were the first researchers to investigate such coupled loops theoretically. They developed a transient 1-D model of the coupled natural circulation loop having toroidal geometry and a point contact heat exchanger section to model the thermal interaction of two adjacent rolls in a Bènard convection layer. Although the model was presented, no validation studies were conducted. The main objective of their study was to perform the stability analysis of the CNCL system.

To model the transient dynamics of a simple toroidal NCL theoretically, Hart [6] had proposed a method utilizing the Fourier series. The method reduced the NCL momentum and energy partial differential equations to a set of three coupled ordinary differential equations. The same method was employed by Davis and Roppo, who restricted the number of Fourier nodes considered for their study to one (as recommended by Hart for toroidal NCL) and a Dirac-Delta function was employed to achieve point thermal coupling between the component loops of the CNCL. An identical temperature profile boundary condition was imposed on both the component NCLs of the system to determine the stability of the point contact toroidal CNCL system.

Davis and Roppo observed that when the temperature differences for both NCLs were equal, stable and steady convective flows were observed over a range of Rayleigh numbers. These flows correspond to parallel and counter configurations at the point heat exchanger section, thus Davis and Roppo had identified the CNCL system as an ideal tool to study the spatial variation in convective flows.

The experimental study of the toroidal CNCL system carried out by Ehrhard [7] validated the theoretical model presented by Davis and Roppo using stability maps. In the experimental study, the volumetric heat generation was used in the heating section and a constant temperature boundary condition for the cooling section. The experimental setup was constructed from glass and the thermal coupling was achieved by using a copper block as the conjugate interface between the glass tubes.

The CNCL systems having a point thermal contact have limited relevance in practical situations, thus a CNCL system constructed from square NCLs, which has a flat plate heat exchanger section that is well suited for industrial applications needs to be studied. A steady-state study employing such a CNCL system was performed by Salazar et al. [8].

The CNCL with flat plate heat exchanger section considered for the theoretical analysis by Salazar et al. was modelled as a simplified version of a Liquid Metal Fast Breeder Reactor (LMFBR). Incorporating several simplifications, a steady state analysis was performed and the coupled velocity and temperature fields were determined. Multiple steady state solutions were reported which corresponded to the clockwise and anticlockwise direction of fluid velocity in each loop of the CNCL. The existence of the multiplicity of solutions was attributed to the nonlinear convective term of the energy equation.

Another work that is of relevance is the experimental study of CNCL system constituting square NCLs performed by Ghoneimy et al. [9]. They conducted a transient study of the CNCL performance employing water as the test fluid in two-phase flow regime at different pressures. The heat exchanger employed in the construction of their CNCL had a complicated structure. Indirect flow measurement techniques were employed by them to characterize the dynamic system behaviour. The experiment was performed to study the capability of the CNCL as a safety mechanism for decay heat removal from the main coolant loop of pressurized water, boiling water and liquid metal nuclear reactors.

Bernal and Vleck [10] presented a 1-D model of a rectangular NCL system, they too employed the Fourier series method used by Hart, but the modelling procedure was modified to suit the rectangular NCL geometry by employing piece-wise functions for representing the geometry and boundary conditions ($+Q''$ and $-Q''$). The validation of the 1D model of NCL for rectangular geometry was carried out by Fichera et al. [3] experimentally.

The CNCL system can be easily visualized as a simplified version of liquid metal fast breeder reactor (LMFBR) and solar domestic water heater (SDWH), both of which employ an intermediate heat exchanger to transfer heat between their loops. The pumping apparatus can be supplanted to make these systems self-actuating and self-sustaining, which is important considering safety parameters, especially in nuclear applications. The LMFBR is illustrated in Fig.1.

From the survey of literature, it is evident that a detailed investigation of the transient behaviour of a CNCL system with emphasis on the practical utilization of the system is still lacking. In this paper the CNCL represented in Fig.2, which has square component NCLs coupled by a flat plate heat exchanger is studied using transient and steady state analysis, with emphasis on the transient system behaviour at low power loads. The current investigation of the CNCL system is confined to low power loads because NCL system exhibits a chaotic behaviour at higher power loads, this characteristic of the NCL has a strong dependence on the initial conditions thus rendering the task of modelling such systems difficult as observed by Rasband [11], Thompson and Stewart [12]. In addition, the determination of the heat transfer coefficient for the chaotic condition is difficult.

The objectives of the present study are:

1. To understand the transient behaviour of a CNCL.
2. Obtain an estimate of the time taken by the system to reach steady state.

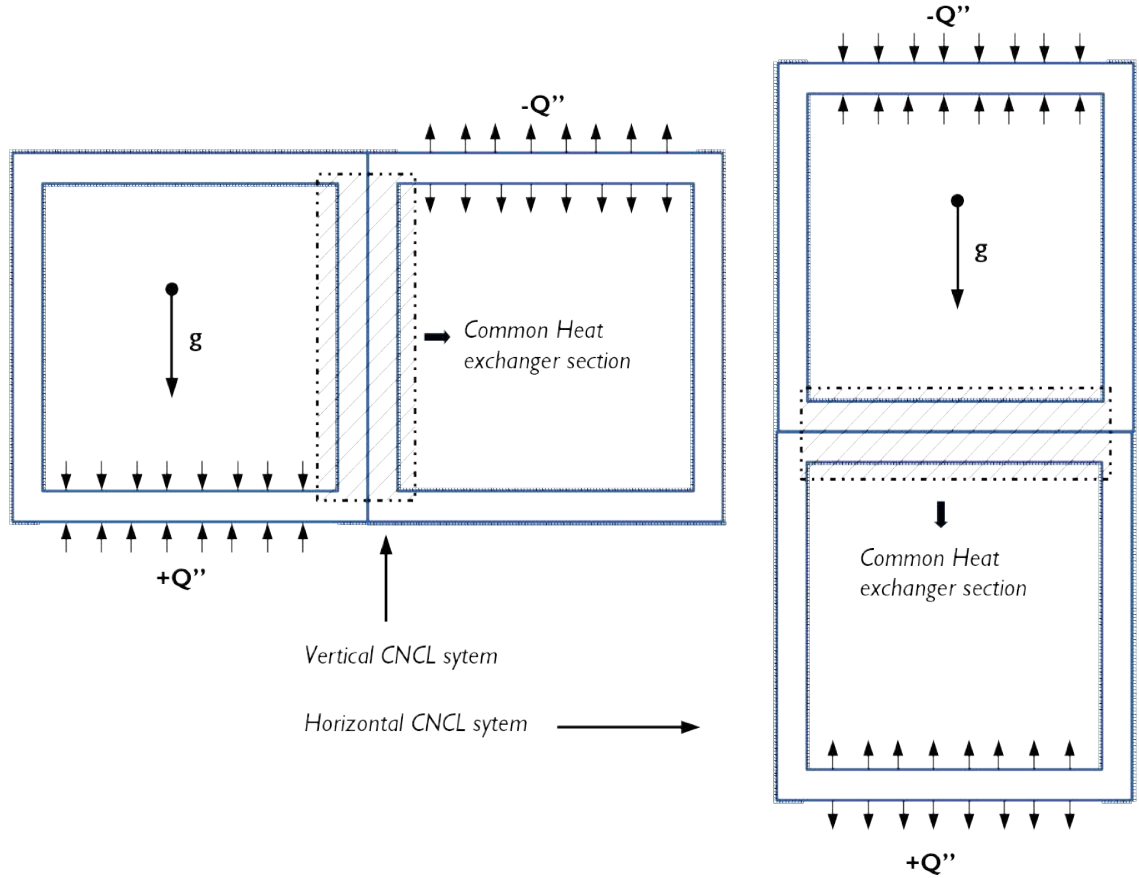

Figure 2: (a) Vertical CNCL and (b) Horizontal CNCL configurations.

3. To determine the effect of initial conditions on the existence of the multiple steady state solutions.
4. To obtain the optimum configuration of the heat source and sink for a rectangular CNCL system such that the heat transfer is maximum and the time required to attain steady state is prolonged or vice-versa.
5. Study of the parallel and counter flow configuration of the heat exchanger section of the CNCL system.
6. Effect of orientation of the heat transfer section of the CNCL.
7. Conduct a complete parametric study of the CNCL system.

To achieve the aforesaid objectives a 1-D single-phase dynamical model of the CNCL composed of square NCLs (referred as the CNCL from hereon) is proposed and its solution is computed using a semi-analytical method. A detailed 3D CFD simulation is undertaken for validating the proposed 1-D model of the CNCL and enhance the understanding of its underlying physics. The paper begins with an introduction to the 1-D model of the CNCL, which includes governing equations, initial and boundary conditions, steady-state analysis and solution methodology. This is followed by a 3-D CFD study and validation of the 1-D CNCL model. Mesh and time step independence studies are included. Further, the results section includes the effects of initial conditions, heat source or sink configuration, power input etc. on the CNCL. The results section also includes a detailed parametric study of the CNCL employing the 1-D CNCL model and its advantages and limitations. Finally, conclusions are drawn from the study.

# Nomenclature

| | |
|---|---|
| $T_{1,Avg}$ | Circuit average temperature of Loop 1 $(K)$ |
| $T_{2,Avg}$ | Circuit average temperature of Loop 2 $(K)$ |
| $x$ | Distance from origin 'O' $(m)$ |
| $\omega_1$ | Fluid Velocity of Loop 1 $(m/s)$ |
| $\omega_2$ | Fluid Velocity of Loop 2 $(m/s)$ |
| $g$ | Gravitational constant $(m/s^2)$ |
| $Q''$ | Heat flux $(W/m^2)$ |
| $h_{1/2}$ | Heat transfer coefficient at the heat exchanger section adjacent to fluid 1 or 2 $(W/m^2K)$ |
| $L$ | Height of the NCL $(m)$ |
| $D_h$ | Hydraulic diameter of both Loop 1 & 2 $(m)$ |
| $U$ | Overall heat transfer coefficient at the heat exchanger section $(W/m^2K)$ |
| $T_0$ | Reference temperature of Loop 1 & 2 $(K)$ |
| $C_p$ | Specific heat capacity $(J/KgK)$ |
| $\omega_{1,ss}$ | Steady state fluid velocity of Loop 1 $(m/s)$ |
| $\omega_{2,ss}$ | Steady state fluid velocity of Loop 2 $(m/s)$ |
| $a$ | Thermal diffusivity $(m^2/s)$ |
| $T_1$ | Temperature of Loop 1 $(K)$ |
| $T_2$ | Temperature of Loop 2 $(K)$ |
| $t$ | Time $(s)$ |
| $L1$ | Width of the NCL $(m)$ |

**Non-dimensional numbers**

| | |
|---|---|
| $K$ | Bend losses coefficient $(K = \Delta P_b end/(\frac{1}{2}\rho \omega_{1,2}^2))$ |
| $f_F$ | Fanning friction factor $(K = \tau_{1,2}/(\frac{1}{2}\rho \omega_{1,2}^2))$ |
| $Nu$ | Nusselt number, $Nu = (h_{1,2}L)/\kappa$ |
| $Re$ | Reynolds number, $Re = (2r(\omega_{1,2}))/\nu$ |
| $Ra$ | Rayleigh number, $Ra = (g\beta(T_{Avg,L1} - T_{Avg,L2})L^3)/(\nu a)$ |

**Greek letters**

| | |
|---|---|
| $\beta$ | Coefficient of thermal expansion $(1/K)$ |
| $\rho$ | Density $(kg/m^3)$ |
| $\nu$ | Kinematic viscosity $(m^2/s)$ |
| $\rho_0$ | Reference Density $(kg/m^3)$ |
| $\kappa$ | Thermal conductivity $(W/(mK))$ |
| $\tau$ | Wall shear stress exerted on fluid $(Pa)$ |

**Subscripts**

| | |
|---|---|
| 0 | Any parameter at time $t = 0\ s$ |
| 1 | Any parameter referring to Loop 1 |
| 2 | Any parameter referring to Loop 2 |
| $ss$ | Any parameter considered at steady state |
| $Avg$ | Average value of parameter |

**Abbreviations**

| | |
|---|---|
| $CFD$ | Computational Fluid Mechanics |
| $CNCL$ | Coupled/Conjugate Natural Circulation Loop |
| $NCL$ | Natural Circulation Loop |
| $LMFBR$ | Liquid Metal Fast Breeder Reactors |
| $SDWH$ | Solar Domestic Water Heater |
| $HX$ | Heat exchanger |

## 2. Mathematical modelling of CNCL system:

Salazar et al. [8] were the first to perform a steady state analysis of the CNCL system with flat plate heat exchanger. The following simplifications were incorporated by Salazar et al. to model the CNCL:

a) The fluids in the primary and secondary loop of the CNCL are assumed to be driven exclusively via natural circulation.

b) The Intermediate Heat Exchanger section of the CNCL is a flat plate heat exchanger.

c) Heat transfer in the heat exchange zone is directly proportional to the local temperature difference between the primary and secondary loop.

d) All thermophyscial properties are assumed to have a constant magnitude.

e) The Boussinesq hypothesis is employed to model the buoyancy forces.

f) 1D NCL momentum and energy equations are used for the study.

g) Water is considered as the operating fluid in the primary and secondary loop, thus adapting the study for Light Water reactors.

h) The axial heat conduction term of the energy equation is disregarded in the analysis performed.

i) Frictional forces are considered to be a linear function of velocity.

The simplifications from (a) to (f) are considered in the present study with the inclusion of the axial heat conduction term and the implementation of a nonlinear function of velocity for the frictional forces. This study extends the analysis by enabling the modelling of transient behaviour of the CNCL system containing different fluids in each of the loops. Both Loop 1 and Loop 2 have identical dimensions and have a square cross-section. The current section examines the 1-D modelling of Vertical CNCL, as the procedure is similar for the Horizontal CNCL configuration. To model the Horizontal CNCL, modifications are made to the piecewise functions representing the geometry and boundary conditions of the loops.

It is to be noted that employing the $-Q''$ on Loop 1 and $+Q''$ on Loop 2 to model the CNCL common heat exchanger section results in a constant average temperature (w.r.t. time) in both loops thus decoupling the two loops. Hence, a convection boundary condition is employed to model the CNCL system.

### 2.1. Governing Equations of the CNCL

The CNCL has two NCLs: Loop 1, Loop 2, which are coupled at the common heat exchanger section. Corners of Loop 1 are denoted using alphabets (A-D) and that of Loop 2 are denoted using alphabets (E-H), as shown in Fig.3. To derive the governing equations of the CNCL system, the CNCL is divided into two parts. Thereby Loop 1 and Loop 2 become two distinct NCLs with the heat exchanger section transforming into the heat sink for the NCL formed by Loop 1 and heat source for the NCL formed by Loop 2. Figure 3 clearly depicts this process and labeling of the system. The objective behind flipping the NCL formed by Loop 1 vertically is to have a common sign for velocity vector ('+ve' in clockwise direction) and have the origin $'O'$ at the left bottom corner. Apart from the sections where the constant heat flux addition or removal condition is enforced, rest of the CNCL is insulated from the atmosphere.

The governing equations of the CNCL system are:

$$\rho_1 \frac{d\omega_1(t)}{dt} + \frac{4\tau_1}{D_h} = \frac{\rho_1 g \beta_1}{2(L+L1)} (\oint (T_1 - T_0) f_1(x) dx) \tag{1}$$

$$\frac{\partial T_1}{\partial t} + \omega_1(t) \left( \frac{\partial T_1}{\partial x} \right) = H_1(x,t) + a_1 \frac{\partial^2 T_1}{\partial x^2} \tag{2}$$

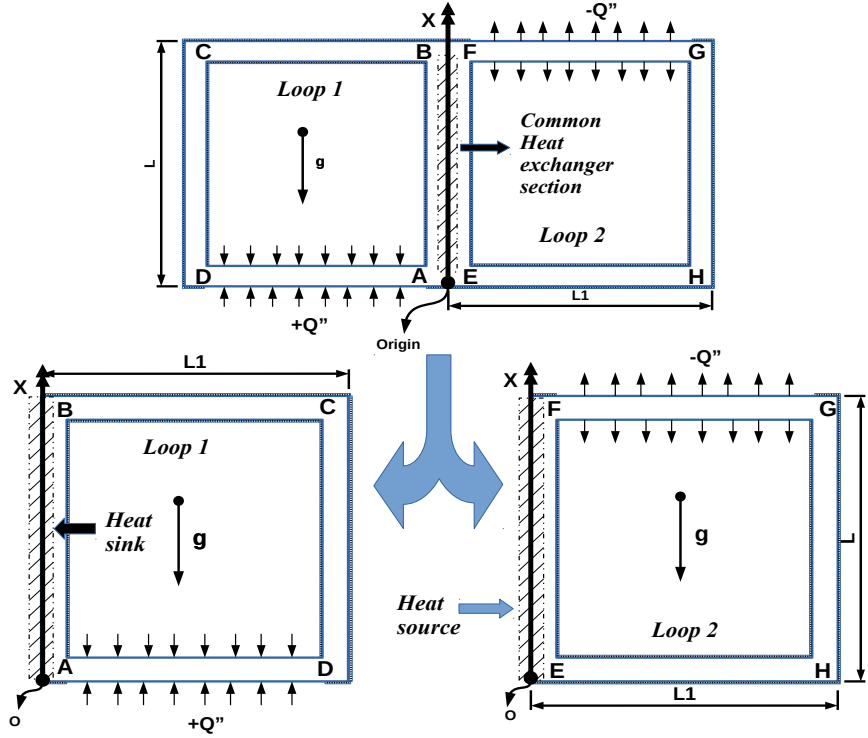

Figure 3: CNCL modelling approach.

$$\rho_2 \frac{d\omega_2(t)}{dt} + \frac{4\tau_2}{D_h} = \frac{\rho_2 g \beta_2}{2(L+L1)} \left( \oint (T_2 - T_0) f_2(x) dx \right) \quad (3)$$

$$\frac{\partial T_2}{\partial t} + \omega_2(t)\left(\frac{\partial T_2}{\partial x}\right) = H_2(x,t) + a_2 \frac{\partial^2 T_2}{\partial x^2} \quad (4)$$

where,

$$\tau_1 = \frac{\rho_1 \omega_1^2 f_{F1}}{2} \quad (5)$$

$$\tau_2 = \frac{\rho_2 \omega_2^2 f_{F2}}{2} \quad (6)$$

Equations 1 and 3 represent the momentum equations of the Loop 1 and Loop 2, respectively. Equations 2 and 4 represent the energy equations of the Loop 1 and Loop 2, respectively. $\tau_1$ and $\tau_2$ represent the shear forces action on the fluid elements of Loop 1 and Loop 2, respectively.

$$H_1(x,t) = \frac{4Q"}{\rho_1 C_{p1} D_h} h_1(x) - \frac{U}{\rho_1 C_{p1} D_h} \lambda(x)(T_1 - T_2) \quad (7)$$

$$H_2(x,t) = \frac{4Q"}{\rho_2 C_{p2} D_h} h_2(x) + \frac{U}{\rho_2 C_{p2} D_h} \lambda(x)(T_1 - T_2) \quad (8)$$

$H_1$ and $H_2$ are functions which signify the net amount of power transferred to each of the constituent NCLs: Loop

1 and Loop 2, respectively.

$$f_1(x) = f_2(x) = \begin{cases} 1 & 0 < x < L \\ 0 & L < x < L + L1 \\ -1 & L + L1 < x < 2L + L1 \\ 0 & 2L + L1 < x < 2(L + L1) \end{cases} \qquad (9)$$

$$\lambda(x) = \begin{cases} 1 & 0 < x < L \\ 0 & L < x < L + L1 \\ 0 & L + L1 < x < 2L + L1 \\ 0 & 2L + L1 < x < 2(L + L1) \end{cases} \qquad (10)$$

$f(x)$ is the function which represents the rectangular geometry of both Loop 1 and Loop 2. $\lambda(x)$ is the function which couples both Loop 1 and Loop 2.

### 2.2. Friction factor Correlations

The frictional forces direct the dynamic behaviour of the CNCL system; hence, it is pivotal to employ proper correlations to determine their magnitude as the flow develops with time. The current study involves flow in a square duct, thus the suitable friction factors for square duct in laminar and turbulent regions are utilized, and where appropriate correlations are unavailable, the hydraulic diameter is employed to evaluate the friction factor. We shall employ the friction factor correlations used by Swapnalee and Vijayan [13]. The general Fanning friction factor correlation is given by equation (11). The values of 'C' and 'd' for various regimes is listed in Table.1

$$f_F = C/Re^d \qquad (11)$$

Table 1: Friction factor constants.

| Flow regime | C | d |
|---|---|---|
| Laminar | 14 | 1 |
| Turbulent | 0.079 | 0.25 |

### 2.3. Initial and boundary conditions of the CNCL system

#### 2.3.1. Initial conditions

The initial condition of the CNCL system is represented by the initial conditions of Loop 1 and Loop 2. The initial condition for the momentum and energy equations of the CNCL system are depicted by equations 12 and 13.

$$w_1(0) = w_0 \ , \ w_2(0) = w_0 \qquad (12)$$

$$T_1(x,0) = T_0 \ , \ T_2(x,0) = T_0 \qquad (13)$$

### 2.3.2. Boundary conditions

$h_1(x)$ is the function which represents the location of the positive heat flux boundary condition of 'Loop 1'. $h_2(x)$ is the function which represents the location of the negative heat flux boundary condition of 'Loop 2'.

$$h_1(x) = \begin{cases} 0 & 0 < x < L \\ 0 & L < x < L + L1 \\ 0 & L + L1 < x < 2L + L1 \\ (4Q'')/(\rho_1 C_{p,1} D_h) & 2L + L1 < x < 2(L + L1) \end{cases} \quad (14)$$

$$h_2(x) = \begin{cases} 0 & 0 < x < L \\ (-4Q'')/(\rho_2 C_{p,2} D_h) & L < x < L + L1 \\ 0 & L + L1 < x < 2L + L1 \\ 0 & 2L + L1 < x < 2(L + L1) \end{cases} \quad (15)$$

### 2.4. ODEs which represent the 1-D single phase CNCL system

To simplify the task of obtaining the solution to the aforementioned partial differential equations (PDE's), we employ the Fourier series expansions of the temperature and boundary conditions to convert the PDE's to ordinary differential equations (ODEs).

$$T_1(x,t) = \sum_{k=-\infty}^{\infty} \alpha_k(t) e^{ik\pi x/(L+L1)} \quad (16)$$

$$T_2(x,t) = \sum_{k=-\infty}^{\infty} \beta_k(t) e^{ik\pi x/(L+L1)} \quad (17)$$

$$h_1(x) = \sum_{k=-\infty}^{\infty} \gamma_k e^{ik\pi x/(L+L1)} \quad (18)$$

$$h_2(x) = \sum_{k=-\infty}^{\infty} \delta_k e^{ik\pi x/(L+L1)} \quad (19)$$

$$\lambda(x) = \sum_{k=-\infty}^{\infty} \zeta_k e^{ik\pi x/(L+L1)} \quad (20)$$

$$f(x) = \sum_{k=-\infty}^{\infty} A_k e^{ik\pi x/(L+L1)} \quad (21)$$

where,

$$\alpha_0(t) = \frac{1}{2(L+L1)} \int_0^{2(L+L1)} T_1(x,t) dx \quad (22)$$

$$\beta_0(t) = \frac{1}{2(L+L1)} \int_0^{2(L+L1)} T_2(x,t) dx \quad (23)$$

for all 't'. In addition $\overline{\alpha_k} = \alpha_{-k}$, the same applies for all aforementioned complex Fourier coefficients.

Substituting the above expressions in equations (1-4) results in the following set of equations after we restrict the number of Fourier nodes to three:

$$\frac{d\omega_1(t)}{dt} + \frac{2C}{D_h}\left(\frac{\nu}{D_h}\right)^d (\omega_1)^{2-d} = g\beta \sum_{n=-3}^{3} \alpha_n A_{-n} \quad (24)$$

$$\frac{d(\alpha_0(t))}{dt}+ = \gamma_0 - \sum_{n=-3}^{3} \frac{U}{\rho_1 C_{p1} D_h} \zeta_{-n}(\alpha_n(t) - \beta_n(t)) \tag{25}$$

$$\frac{d(\alpha_1(t))}{dt} + \frac{i\pi}{L+L1}w_1(t)\alpha_1(t) = \gamma_1 - \frac{a\alpha_1(t)\pi^2}{(L+L1)^2} - \sum_{n=-3}^{3} \frac{U}{\rho_1 C_{p1} D_h} \zeta_{1-n}(\alpha_n(t) - \beta_n(t)) \tag{26}$$

$$\frac{d(\alpha_2(t))}{dt} + \frac{2i\pi}{L+L1}w_1(t)\alpha_2(t) = \gamma_2 - \frac{4a\alpha_2(t)\pi^2}{(L+L1)^2} - \sum_{n=-3}^{3} \frac{U}{\rho_1 C_{p1} D_h} \zeta_{2-n}(\alpha_n(t) - \beta_n(t)) \tag{27}$$

$$\frac{d(\alpha_3(t))}{dt} + \frac{3i\pi}{L+L1}w_1(t)\alpha_3(t) = \gamma_3 - \frac{9a\alpha_3(t)\pi^2}{(L+L1)^2} - \sum_{n=-3}^{3} \frac{U}{\rho_1 C_{p1} D_h} \zeta_{3-n}(\alpha_n(t) - \beta_n(t)) \tag{28}$$

$$\frac{d\omega_2(t)}{dt} + \frac{2C}{D_h}\left(\frac{\nu}{D_h}\right)^d (\omega_2)^{2-d} = g\beta \sum_{n=-3}^{3} \beta_n A_{-n} \tag{29}$$

$$\frac{d(\beta_0(t))}{dt} = \delta_0 + \sum_{n=-3}^{3} \frac{U}{\rho_2 C_{p2} D_h} \zeta_{-n}(\alpha_n(t) - \beta_n(t)) \tag{30}$$

$$\frac{d(\beta_1(t))}{dt} + \frac{i\pi}{L+L1}w_2(t)\beta_1(t) = \delta_1 - \frac{a\beta_1(t)\pi^2}{(L+L1)^2} + \sum_{n=-3}^{3} \frac{U}{\rho_2 C_{p2} D_h} \zeta_{1-n}(\alpha_n(t) - \beta_n(t)) \tag{31}$$

$$\frac{d(\beta_2(t))}{dt} + \frac{2i\pi}{L+L1}w_2(t)\beta_2(t) = \delta_2 - \frac{4a\beta_2(t)\pi^2}{(L+L1)^2} + \sum_{n=-3}^{3} \frac{U}{\rho_2 C_{p2} D_h} \zeta_{2-n}(\alpha_n(t) - \beta_n(t)) \tag{32}$$

$$\frac{d(\beta_3(t))}{dt} + \frac{3i\pi}{L+L1}w_2(t)\beta_3(t) = \delta_3 - \frac{9a\beta_3(t)\pi^2}{(L+L1)^2} + \sum_{n=-3}^{3} \frac{U}{\rho_2 C_{p2} D_h} \zeta_{3-n}(\alpha_n(t) - \beta_n(t)) \tag{33}$$

The equations [26-28] and [31-33] are separated into real and imaginary parts, thereby obtaining a set of 16 ordinary differential equations which need to be solved simultaneously to obtain $T_1(x,t)$ and $T_2(x,t)$. The Fourier coefficients are plugged into equation [34,35] to obtain temperature distributions in Loop 1 and Loop 2.

$$T_1(x,t) = \sum_{k=-3}^{3} \alpha_k(t) e^{ik\pi x/(L+L1)} \tag{34}$$

$$T_2(x,t) = \sum_{k=-3}^{3} \beta_k(t) e^{ik\pi x/(L+L1)} \tag{35}$$

#### 2.4.1. Evaluation of the steady state heat transfer coefficient

The heat transfer coefficient $U$ is evaluated by employing the fact that the net heat transfer to both Loops 1 and 2 is zero at steady state. Thus applying the energy conservation for Loop 1 we get,

$$4Q"L = \int_0^L U\Big(T_{1,ss}(x) - T_{2,ss}(x)\Big) dx \tag{36}$$

$$U = \frac{4Q"L}{\int_0^L \Big(T_{1,ss}(x) - T_{2,ss}(x)\Big) dx} \tag{37}$$

Since the current paper deals with fluids having high thermal conductivity, the LMTD relation cannot be employed to calculate the heat transfer coefficient as the axial conduction along the flow direction will be significant.

## 2.5. Initial conditions of the ODE system

To obtain the transient behaviour of the CNCL system we need to integrate the ODEs represented by equations (23-32) for which the initial conditions for each of the Fourier nodes need to be determined. The initial conditions of the ODEs are derived from the initial conditions of the momentum and energy equation for both Loop 1 and Loop 2. Since we are limiting ourselves to three nodes, we split the loop into three equal parts and apply the initial conditions mentioned in equations (12) and (13). This results in a matrix of 4 equations and 4 unknowns for each of the loops, as represented below.

$$\begin{bmatrix} 1 & 1 & 1 & 1 \\ 1 & e^{i2\pi/3} & e^{i4\pi/3} & e^{i6\pi/3} \\ 1 & e^{i4\pi/3} & e^{i8\pi/3} & e^{i12\pi/3} \\ 1 & e^{i6\pi/3} & e^{i12\pi/3} & e^{i18\pi/3} \end{bmatrix} \times \begin{bmatrix} \alpha_0 \\ \alpha_1 \\ \alpha_2 \\ \alpha_3 \end{bmatrix} = \begin{bmatrix} T_0 \\ T_0 \\ T_0 \\ T_0 \end{bmatrix}$$

$$\begin{bmatrix} 1 & 1 & 1 & 1 \\ 1 & e^{i2\pi/3} & e^{i4\pi/3} & e^{i6\pi/3} \\ 1 & e^{i4\pi/3} & e^{i8\pi/3} & e^{i12\pi/3} \\ 1 & e^{i6\pi/3} & e^{i12\pi/3} & e^{i18\pi/3} \end{bmatrix} \times \begin{bmatrix} \beta_0 \\ \beta_1 \\ \beta_2 \\ \beta_3 \end{bmatrix} = \begin{bmatrix} T_0 \\ T_0 \\ T_0 \\ T_0 \end{bmatrix}$$

Solving the matrix equations yields the initial value of the Fourier nodes of Temperature. The solution is $\alpha_0 = \beta_0 = T_0$, whilst the values at all other nodes is zero.

## 2.6. Numerical Integration

The set of ODE's obtained by using the Fourier series expansion cannot be solved analytically. Thus to obtain the dynamics of the system we need to use numerical integration techniques such as the Runge-Kutta method which has been employed for the current study using the MATLAB solver 'ode45'.

## 3. 3-D CFD study of a single phase CNCL system

A detailed CFD investigation of the single phase CNCL system has been conducted to explore the physics and dynamic characteristics. The main objective of this CFD study is to validate the 1-D CNCL model rigorously and probe its efficacy. The 3-D CFD study has been performed to study the single phase CNCL system for the set of conditions specified in Table.2.

Table 2: Cases considered for the CFD simulation.

| Case | Loop 1 | Loop 2 | CNCL Orientation | Description |
|---|---|---|---|---|
| CNCL-(a) | FF1 | FF1 | Vertical | Same fluid in both loops |
| CNCL-(b) | FF1 | FF2 | Vertical | Different fluids in the loops |
| CNCL-(c) | FF1 | FF1 | Horizontal | Same fluid in both loops and parallel flow configuration |
| CNCL-(d) | FF1 | FF1 | Horizontal | Same fluid in both loops and counter flow configuration |

The developed 1-D model has been validated with the aforementioned CFD cases to display its robustness for use in prediction of the behaviour of CNCL system. The CFD investigation of the CNCL system has been done using ANSYS Fluent, which is a commercial finite volume software. The pre-processing, processing and post-processing were performed employing the same software. FF1 and FF2 listed in Table.2 are the fluids used in this study, the thermophysical properties of which are elaborated in section 3.3.1.

*3.1. Geometry*

The geometry comprises of two parts: Loop 1 and Loop 2 which are coupled at the common heat exchanger section (Flat plate heat exchanger). Loop 1 & 2 are square NCLs with a square cross-sectional area. Figure4 represents the CNCL geometry considered for the study. The 90 degree bends are chamfered to reduce the bend losses encountered by the flow. Apart from the cooling and heating sections, rest of the loop is completely insulated. The dimensions of the geometry considered for the study is listed in Table.3.

Table 3: Dimensions of the 3-D geometry

| Parameter | Description | Dimensions |
|---|---|---|
| L | Height/Width of the CNCL | 1 $m$ |
| L1 | Height/Width of the CNCL | 1 $m$ |
| $D_h$ | Length of square cross section | 0.04 $m$ |

*3.2. Meshing*

The ANSYS Design Modeler was used to generate a structured 3-D mesh from the geometry, as shown in Fig.5a. The Multi-zone method was used along with the body sizing option to generate a hex-dominant high quality mesh. Average element quality of 0.98 was obtained for all the meshes considered in the present study, Figure 5b shows a bar graph of the element quality of the grid elements which indicates that the mesh is structured through the entire volume. The geometry was named to set the boundary conditions for the CNCL system.

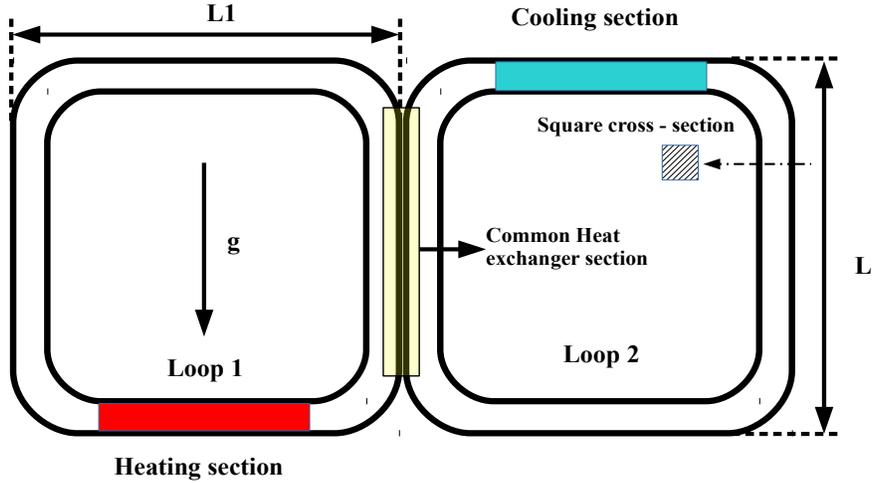

Figure 4: CNCL geometry made of square NCLs Loops 1 and 2 with a square cross section considered for CFD study. The heating and cooling sections, gravity are shown for representative purposes, actual position or orientation depends on the case being investigated.

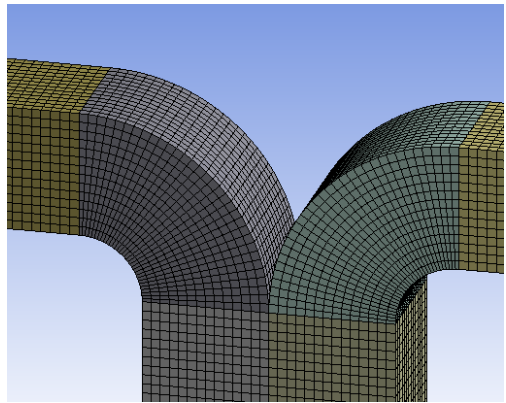

(a) 3-D structured hex-dominant mesh of the CNCL system.

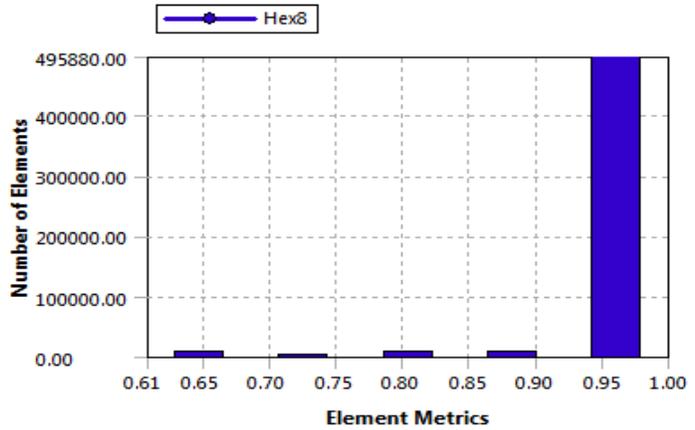

(b) Plot of No of elements vs Element quality of the structured grid used for the CFD study. Element metric of $'1'$ indicates a highly structured mesh.

Figure 5: Mesh characteristics.

*3.3. Case setup*

A transient pressure based solver was used to simulate the transient dynamics of the CNCL system considered. The energy and momentum equation were solved simultaneously on the 3-D grid to obtain the numerical solution at each time step. The laminar flow model was used for the current simulation with the gravity set to 9.81 $m/s^2$.

The CFD study performed by Kudariyawar et al.[14] clearly indicates that for low power loads such as those employed in the current work, the laminar flow model is appropriate for the CFD study and accurately predicts the transient behaviour of the parameters considered.

The Boussinesq approximation has been used to model the buoyancy forces. The operating temperature at start up is set to 300 $K$ and the operating pressure is set to 1 $atm$.

*3.3.1. Fluid used for the CFD study*

It is observed that, the fluids which are normally used have very low values of thermal diffusivity, which becomes an issue when one tries to numerically solve the coupled momentum and energy equations of buoyancy driven flows. The low values of thermal diffusivity impose the selection of an extremely fine mesh and a small time step to get a numerically stable CFD solution. Thus if we employ common fluids such as water, for a given heat input they might require about 20 $min$ to reach the steady state ([3]). Besides, the extremely fine mesh and minute time step requirements make the computational process extremely expensive both in terms of memory and in terms of the clock time if we simulate until the fluid reaches a steady state. Since the main objective is to validate the 1-D semi-analytical model developed we employ a fictitious fluid (a fluid with assumed thermophysical properties) to relax the mesh and time step conditions for numerical stability and hasten the time required to reach the steady state. The fluid chosen for simulation must be irrelevant as long as the underlying physics is the same. Since fictitious fluid utilized has a high thermal diffusivity, it will also enable us to validate its implementation in the 1-D semi-analytical model developed. The thermophysical properties of the fictitious fluids are listed in Table.4.

Table 4: Fluids used for the CFD study.

| Property | Fictitious fluid 1 (FF1) | Fictitious fluid 2 (FF2) |
| --- | --- | --- |
| $\rho_0$ | 70 $kg/m^3$ | 50 $kg/m^3$ |
| $C_p$ | 100 $J/kgK$ | 70 $J/kgK$ |
| $\beta$ | 0.01 $1/K$ | 0.08 $1/K$ |
| $a$ | 0.4 $m^2/s$ | 0.8 $m^2/s$ |
| $\mu$ | 0.0007 $kg/ms$ | 0.005 $kg/ms$ |

*3.3.2. Initial and boundary conditions*

At time $t = 0$ the operating temperature is $T_0$ and the operating pressure is $P_0$. Apart from the heating, cooling and coupled heat exchanging section, rest of the loop is completely insulated. We apply a constant heat flux $Q''$ at the heating section and the corresponding negative value at the cooling section. The coupling that was done in the geometry section results in the formation of a shadow for the coupled wall region. Table.5 lists the initial and boundary conditions of the CFD system used for the current simulation for all the conditions (a-d) mentioned in Table.2.

Table 5: Initial and boundary conditions

| Parameter | Description | Values |
| --- | --- | --- |
| $T_0$ | Temperature at time $t = 0$ | 300 $K$ |
| $P_0$ | Pressure at time $t = 0$ | 1 $atm$ |
| $Q''$ | Constant heat flux supplied or extracted | 2000 $W/m^2$ |
| $g$ | Gravitational constant | 9.81 $m/s^2$ |

*3.3.3. Solver settings and discretization schemes*

The PISO scheme was used for pressure velocity coupling. Least squares cell based scheme was used for the gradient spatial discretization. A second order scheme for Pressure discretization, and second order upwind scheme for the momentum and energy discretization was used respectively. Default solution controls with residuals of order $10^{-3}$ were used and the average velocity and temperature were used as parameters for judging the convergence. The velocity field was initialized to zero and the temperature in the entire domain at time $t = 0$ $s$ was set to 300 K. A fixed time stepping method was used and the number of iterations per time step was set to 200.

### 3.4. Grid independence and time step independence tests

To confirm the reliability of the results obtained, grid and time step independence tests were performed to ensure that the spatial and temporal discretization errors have a minimal impact on the physics of the study. Figures 6a and 6b represents the time step and grid independence results, respectively. The tests are considered to be satisfactory when percentage error is less than 5%. From the tests it is observed that a time step of 1 $s$ and no of elements = 520968 to be ideal for CFD study for the case when both the loops of the CNCL are filled with FF1. The grid and time step studies were performed for all the cases used for validation with the 1-D semi-analytical model.

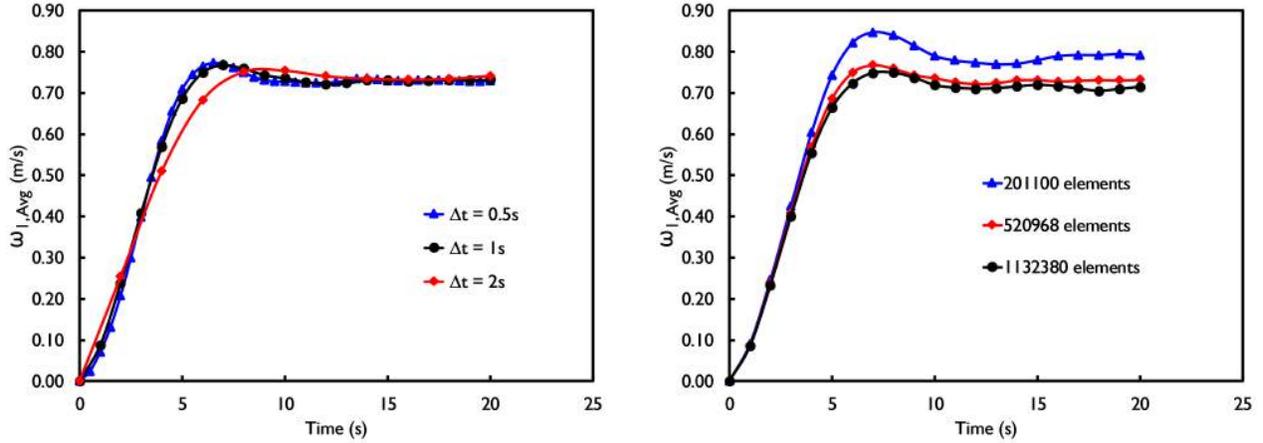

(a) Time step independence test performed with a grid of 520968 elements.

(b) Grid independence test carried out with a time step of size $\Delta t = 1\ s$.

Figure 6: Reliability evaluation of the CFD simulation employing grid and time step independence tests.

## 4. CFD results and analysis

This section provides a brief discussion of the 3-D CFD simulation of the horizontal and vertical CNCL systems. The heat transfer and transient dynamics observed in the CNCL systems are governed by coupled velocity and temperature fields of each of its constituent NCLs as observed in equations (1-4) along with the thermal coupling in the heat exchanger section of the CNCL. Thus the velocity and temperature distribution in both Loop 1 and 2 needs to be studied to characterize a CNCL system. The velocity and temperature contours at steady state are illustrated in Fig.7.

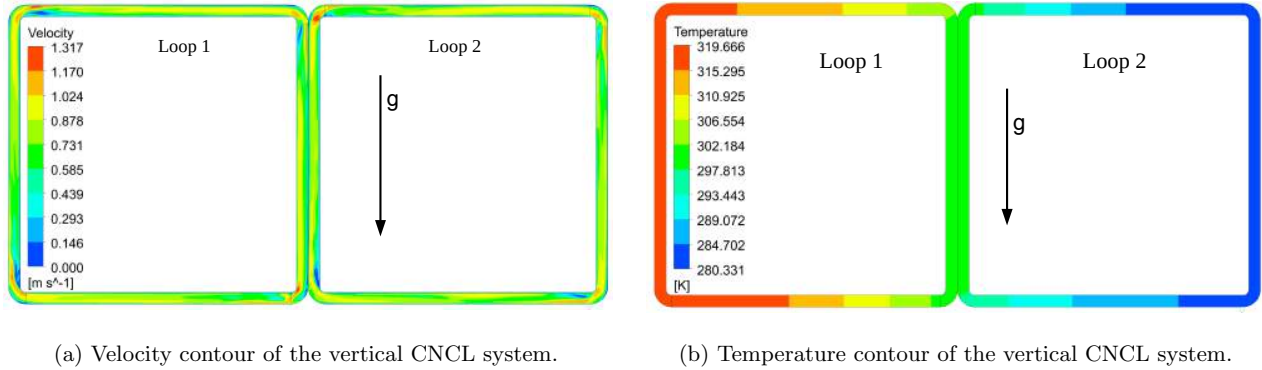

(a) Velocity contour of the vertical CNCL system.  (b) Temperature contour of the vertical CNCL system.

Figure 7: Steady state contours of vertical considered for study with both the loops containing FF1.

From Fig.7a we observe that the velocity of the loop is uniform across the entire domain except near the bends of the CNCL where the velocity magnitude is relatively lower. This indicates that the 90 ° elbow bends have a significant impact on the flow. Figure 7b indicates that temperature drop or rise occurs only along the direction of flow, with very little radial variation in the temperature profile. This is mainly due to the high thermal conductivity of the fluid considered for the study, but the same behaviour may be observed in fluids of lower thermal conductivity because of the churning effect of the vortices that are created at the bends.

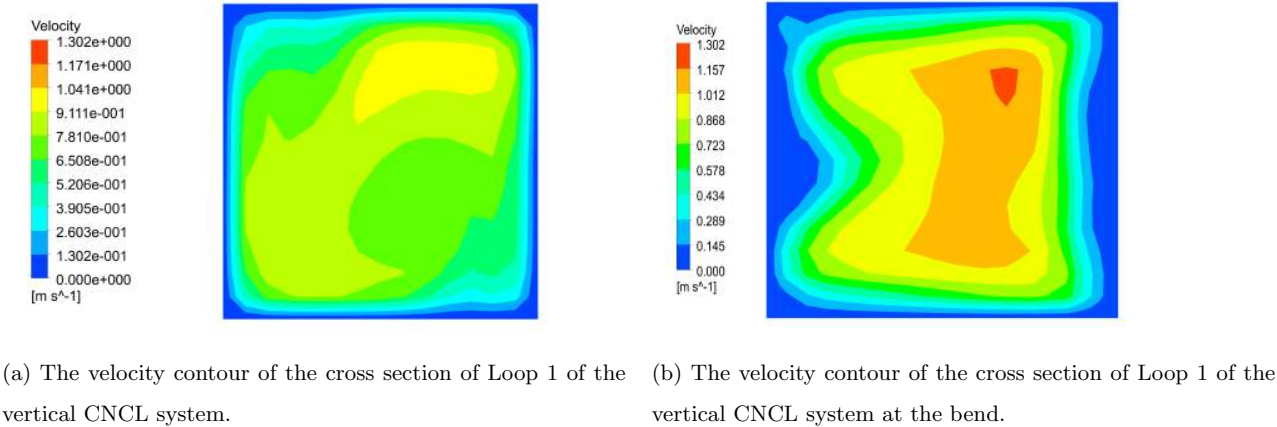

(a) The velocity contour of the cross section of Loop 1 of the vertical CNCL system.  (b) The velocity contour of the cross section of Loop 1 of the vertical CNCL system at the bend.

Figure 8: Steady state contours of vertical considered for study at steady state with both the loops containing FF1 at the cross sectional plane.

The velocity contours of the cross section of Loop 1 of the horizontal CNCL system is illustrated in Fig.8. From Fig.8a we observe that the velocity is zero at the walls and gradually becomes higher in magnitude at the centre, while the peculiar distribution of velocity observed in Fig,8b is because of the centrifugal forces acting upon on the fluid at the 90° smooth elbow bend.

The above-mentioned characteristics are common to all the CNCL cases studied. Hereafter, only the parameters of paramount importance to the study are presented.

### 4.1. Transient dynamics of the CNCL system

Average velocity and average temperature are utilised as parameters to compare the 3-D CFD case with the 1-D semi analytical model. These parameters are averaged over the volume of the respective NCLs, which make up the CNCL system. Figure 9 represents the CFD data obtained from the 3-D simulation of the CNCL system.

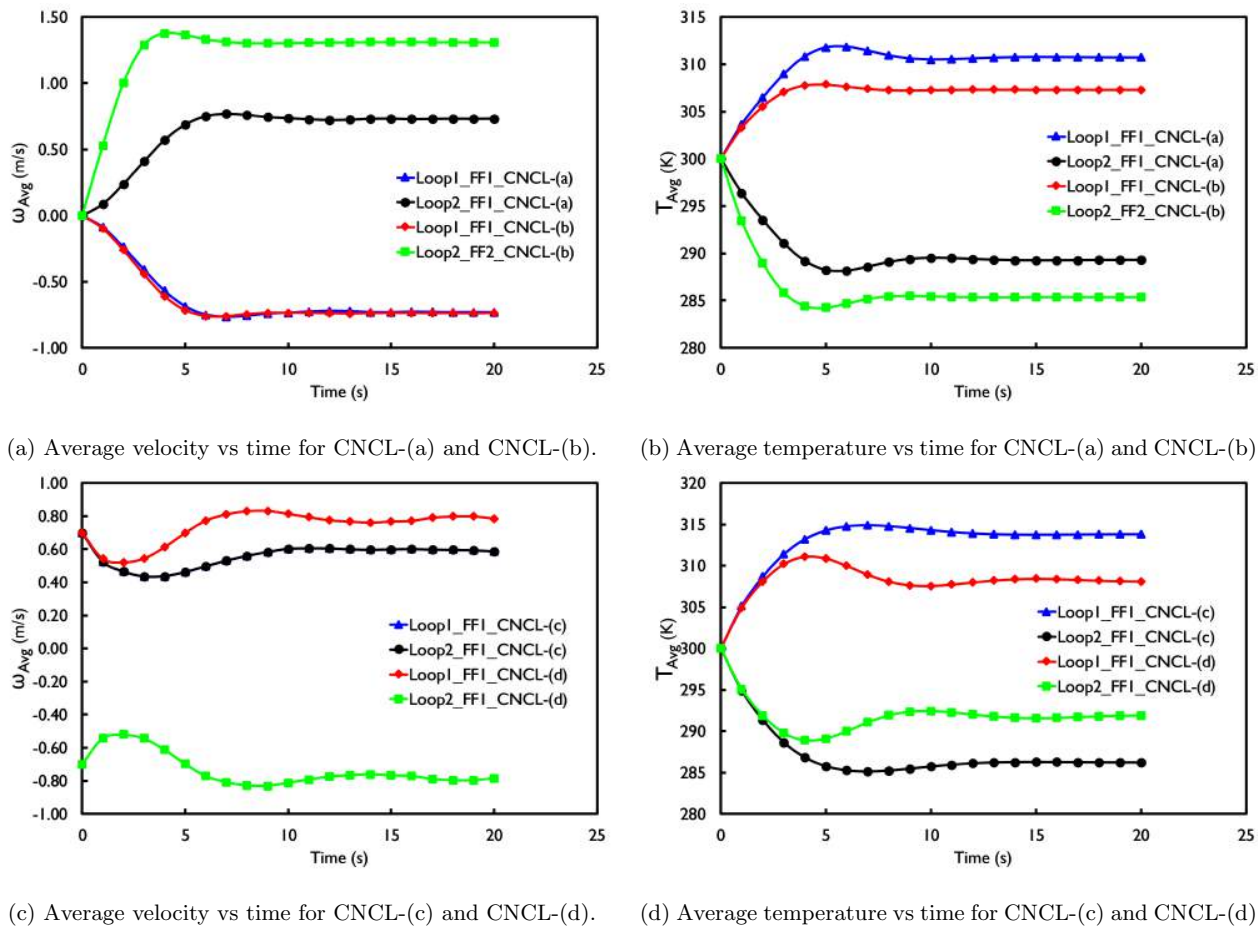

(a) Average velocity vs time for CNCL-(a) and CNCL-(b).  (b) Average temperature vs time for CNCL-(a) and CNCL-(b).

(c) Average velocity vs time for CNCL-(c) and CNCL-(d).  (d) Average temperature vs time for CNCL-(c) and CNCL-(d).

Figure 9: Transient volume averaged plots of velocity and temperature of the vertical and horizontal CNCL system obtained using CFD. The legend of the graphs is interpreted in the following manner: '$Loop1-FF1-CNCL-(a)$' indicates that for CNCL-(a) which represents a vertical CNCL with the component 'Loop 1' filled with fluid FF1.

From Fig.9 the following observations can be made:

1. The introduction of a different fluid in one of the loops( as in case of CNCL-(b)) does not seem to affect the transient average velocity plot of the other loop, but the transient temperature plot is influenced significantly as observed from Fig.9a and Fig.9b which compare CNCL-(a) and CNCL-(b). This is because of the fact that the velocity field is only influenced by the frictional forces at the steady state for the cases compared.

2. From Fig.9b we can observe that when both loops contain the same fluid the average transient temperature plot is symmetric about $T_0$ (which is the initial value of the average temperature of the constituent loops of the CNCL system), but when a different fluid (as in case of CNCL-(b)) is introduced in one of the loops this symmetry is lost.

3. From Fig.9c(horizontal CNCL system) we observe that both the loops are initiated with a certain velocity, depending on the direction of velocity in each of the loops the flow can either be in counter-flow(CNCL-(d)) or parallel-flow configuration(CNCL-(c)). The counter-flow configuration has a greater velocity magnitude in comparison with the parallel-flow configuration for the horizontal CNCL system.

4. From Fig.9d we observe that the average temperature of the counter-flow configuration is lower in magnitude than the parallel-flow configuration. This is because of the uniform temperature drop along the length of the heat exchanger section for the counter-flow arrangement in contrast with the parallel-flow arrangement.

### 4.2. Transient behaviour of the overall heat transfer coefficient of the CNCL system

Figure10 represents the transient nature of the overall heat transfer coefficient ($U$) at the common heat exchanger section. Initially the heat transfer drops from infinity for cases- CNCL-(a) to CNCL-(d), which indicates the growth of the thermal boundary layer, after the boundary layer thickness approaches a constant magnitude no variation in $U$ is observed. The Horizontal CNCL system in counter-flow configuration represented by CNCL-(d) has the largest magnitude of heat transfer coefficient at the steady state.

For the vertical CNCL system represented by cases: CNCL-(a) and CNCL-(b), the heat transfer is determined by the buoyancy forces generated at the heat exchange section, thus the heat transfer coefficient correlations for free convection need to be employed to model this case.

For the Horizontal CNCL system represented by cases: CNCL-(c) and CNCL-(d), the heat transfer is determined by the buoyancy forces generated in the insulated sections of the loop, which cause the flow to occur in the heat exchanger section thus the forced convection heat transfer coefficients need to be employed to model this case.

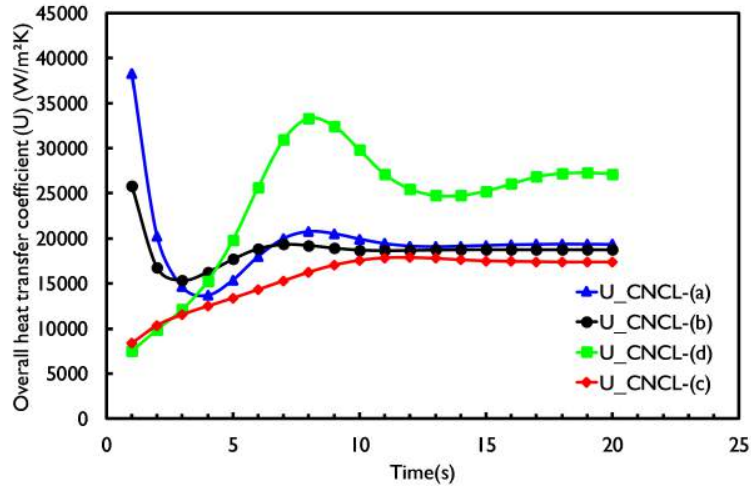

Figure 10: CFD study of overall heat transfer coefficient ($U$) variation with time.

### 4.3. Effect of bends on the friction factor of the CNCL system

The 3-D CFD study of the CNCL system reveals that vortices and eddies are generated as the flow field develops with time, and these vortices have a significant impact on the energy distribution of the CNCL system. It needs to be stressed here that the vortices generated are not because of turbulence but because of the geometry of the CNCL and the effect of bends. For fluids considered for the study, the friction factor is significantly affected by the bends. This is best visualised by doing a power law data-fit of the CFD data and comparing the coefficients of the correlation with the friction factors used in the study. The transient shear stress versus velocity plot is represented in Fig.11 and the theoretical expression for wall shear is given by equation-38.

$$\tau_1 = \frac{1}{2}\rho_0 C \left(\frac{\nu}{D_h}\right)^d \omega_1^{2-d} \qquad (38)$$

The values of 'C' and 'd' are computed using equation-38 and a comparison with the laminar and turbulence parameters used for the present study is shown in Table.6. The values of the parameter are closer to the laminar flow values and the deviation in magnitude is due to the effect of bends, which indicates that bend losses cannot be neglected. Figure12 represents the vortices formed at the elbows of Loop 1 and Loop 2 clearly indicated by the streamlines and velocity vectors.

Table 6: Friction factor constants.

| Flow regime | C | d |
|---|---|---|
| Laminar | 14 | 1 |
| Turbulent | 0.079 | 0.25 |
| CFD datafit | 11.6474 | 0.8509 |

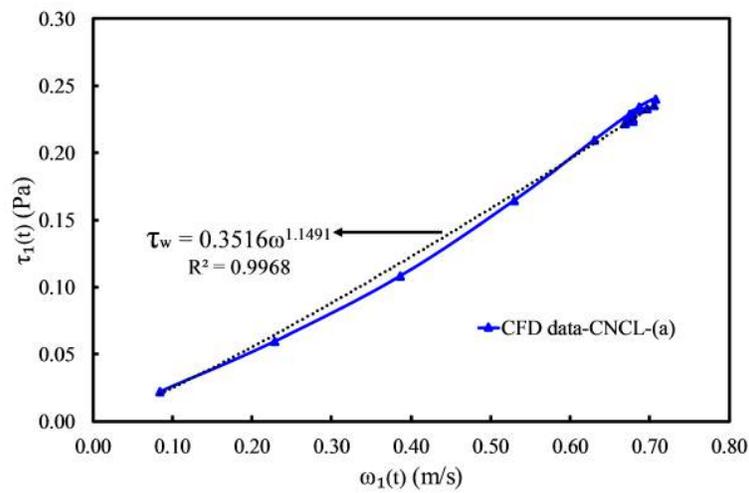

Figure 11: Transient shear stress vs velocity plot with a power law data fit.

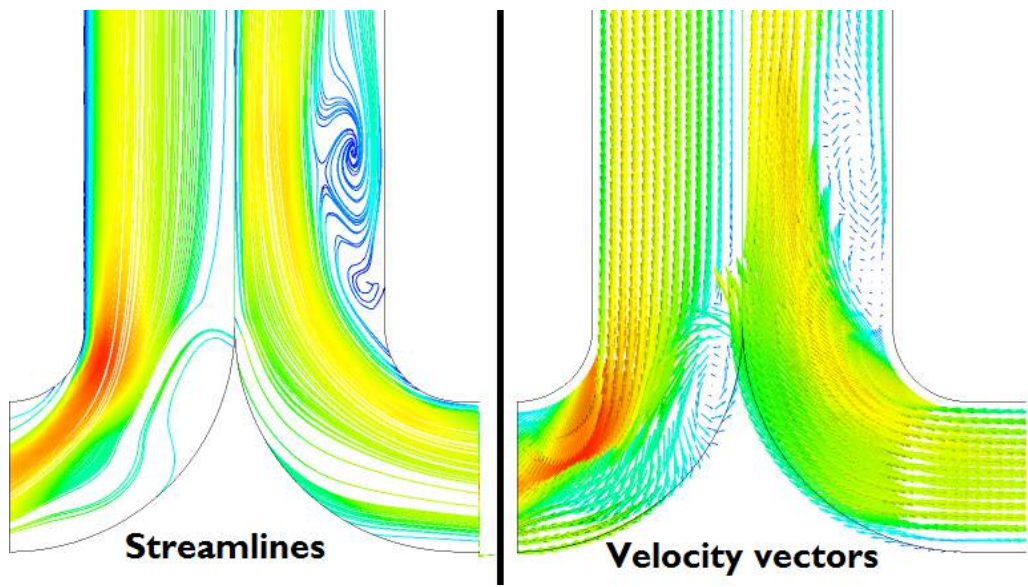

Figure 12: Vortices formed near the bends of the CNCL system clearly indicated by the streamlines and velocity contours.

*4.4. Effect of different cooler and heater configurations on the vertical and horizontal CNCL system*

The transient dynamics of the CNCL system is determined significantly by the orientation of the heater and cooler. Thus to find the most suitable heater and cooler orientation for practical applications, all the possible configurations as shown in Fig.13 are explored for Vertical CNCL, Horizontal CNCL with an initial temperature of 300K and initial velocity of 0.7 m/s to get counter or parallel flow configuration in the heat exchanger section.

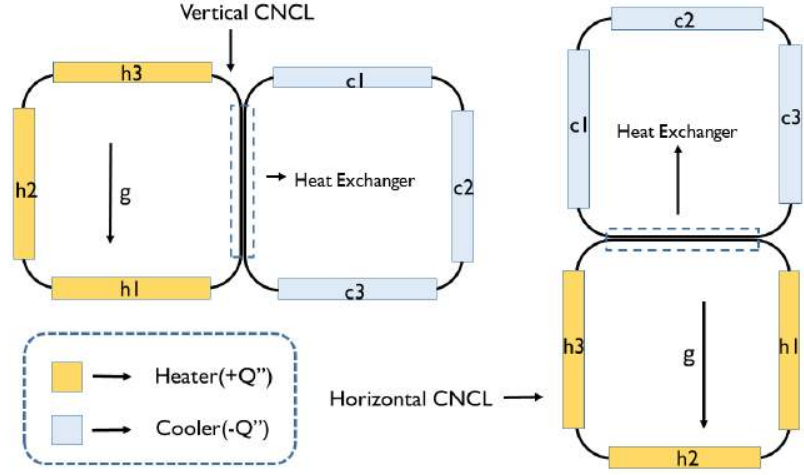

Figure 13: Heater and cooler configurations used for the present study.

In this study, we are solely interested in studying the effect of one heater and cooler per CNCL system. The configuration 'Vertical CNCL h1c1' refers to a vertical CNCL system where, only heater-'h1' and cooler-'c1' are switched on ($Q'' = 2000\ W/m^2$) and the remaining heaters('h2' and 'h3') and coolers ('c2' and 'c3') are set to $Q'' = 0$ $W/m^2$. A 3-D steady state analysis was conducted using CFD software (Fluent) to perform this study.

This study enables us to discern the trend of average velocity, temperature and the heat transfer coefficient at steady state for CNCL systems operating with a single fluid that has similar thermophysical characteristics as FF1 or FF2. This implies that we shall observe a similar trend when liquid metals are used to carry out the study. Table.7 shows the average fluid velocities and temperatures obtained for different configurations using the steady state 3-D CFD simulations.

Some observations which can be drawn from data in Table.7 are:

1. For CNCL-(a) the maximum heat transfer coefficient ($U$) corresponds to 'h2c2' configuration, in this configuration the heater, cooler and heat exchanger sections are all vertically aligned with respect to gravity and the distance between the heater and cooler centres is maximum.

2. For CNCL-(c) the minimum heat transfer coefficient (U) corresponds to configuration 'h2c2', where the heater, cooler and the heat exchanger are horizontal with respect to gravity.

3. For CNCL-(d) the minimum heat transfer coefficient (U) corresponds to configurations'h1c3', 'h3c1' and 'h1c2', these configurations also have low average fluid velocities. This supports the claim that for a horizontal CNCL system the average heat transfer coefficient is a function of Reynolds number.

Table 7: Effect of heater and cooler configuration on the CNCL system.

| Case | Configuration | Average fluid velocity (m/s) | Average fluid temperature (K) | U ($W/m^2K$) |
|---|---|---|---|---|
| CNCL-(a) | h1c1 | 0.676059 | 309.793 | 18315.5 |
| | h1c2 | 0.683655 | 309.587 | 21943.3 |
| | h1c3 | 0.670582 | 308.075 | 20970.1 |
| | h2c1 | 0.772252 | 311.107 | 21621.8 |
| | h2c2 | 0.730812 | 307.58 | 26750.9 |
| | h2c3 | 0.712707 | 307.496 | 24508.1 |
| | h3c1 | 0.400785 | 306.49 | 20742 |
| | h3c2 | 0.36444 | 304.745 | 24492.7 |
| | h3c3 | 0.344624 | 303.767 | 22189.5 |
| CNCL-(c) | h1c1 | 0.749803 | 301.168 | 28895.1 |
| | h1c2 | 0.721985 | 302.856 | 28024.4 |
| | h1c3 | 0.584 | 305.6 | 18266.2 |
| | h2c1 | 0.769922 | 300.33 | 28044.9 |
| | h2c2 | 0.6164 | 314.344 | 18167.2 |
| | h2c3 | 0.774498 | 302.147 | 28048.4 |
| | h3c1 | 0.584 | 305.6 | 18266.2 |
| | h3c2 | 0.619528 | 307.632 | 18226.1 |
| | h3c3 | 0.749803 | 301.168 | 28895.1 |
| CNCL-(d) | h1c1 | 0.768 | 300.7 | 29377 |
| | h1c2 | 0.524508 | 316.985 | 18146.2 |
| | h1c3 | 0.498114 | 311.769 | 18233.3 |
| | h2c1 | 0.779647 | 305.088 | 28356.7 |
| | h2c2 | 0.774482 | 306.842 | 27497.9 |
| | h2c3 | 0.779647 | 305.088 | 28356.7 |
| | h3c1 | 0.498144 | 311.769 | 18233.3 |
| | h3c2 | 0.742514 | 302.815 | 28346.5 |
| | h3c3 | 0.768 | 300.7 | 29377 |

## 5. Validation of the 1-D model with the 3-D CFD numerical simulation

An exhaustive 3-D CFD study was performed to properly understand the dynamics and physics of the CNCL system for all the cases mentioned in Table.2. We shall now employ the CFD results to validate the 1-D semi analytical model developed. The 3-D numerical simulations were performed using a commercial software and are computationally expensive both in-terms of the data they generate and the amount of time required to simulate the transient dynamics. Thus, the main motivation of developing a 1-D model is to remove the dependence on software, reduce the amount of data generated and get a proper insight into the physics of the system while expending minimum amount of time in the process. Table.8 illustrates a comparative study on the time taken for the 3-D CFD and 1-D modelling approaches.

Table 8: Comparison of the time required for each of the parameters for a 3-D CFD case vs 1-D semi analytical model.

| Parameter under consideration | 3-D CFD Study | 1-D semi analytical model |
|---|---|---|
| Geometry creation | minimal | not required |
| Structured mesh generation | substantial | not required |
| Case setup | substantial | minimal |
| Simulation | significantly large | minimal |
| Post processing | significantly large | substantial |
| Grid and time step indpendence | significantly large | minimal |

*5.1. Heat transfer coefficient of the CNCL heat exchanger*

As mentioned previously (section.4.2), to model the heat transfer coefficient in the heat exchanger section we need to employ a piece-wise correlation to represent the transient variation of the heat transfer coefficient for vertical and horizontal CNCL systems. The transient behaviour of the heat transfer coefficient from time $t = 0\ s$ as observed from Fig.10 can be attributed to the thermal boundary layer growth, which is solely determined by conduction. The heat transfer coefficient in the conduction regime is calculated using equation-39.

$$h_{1/2,conduction}(t) = \frac{\kappa}{\sqrt{\pi a t}} \tag{39}$$

For the vertical CNCL system the heat transfer is driven directly by the buoyancy forces in the heat exchanger section, thus free convection correlations are employed to model such systems, the heat transfer coefficient is determined in the most general format using equation-40.

$$h_{1/2,VerticalHX}(t) = C_1 Ra(t)^{1/4} \tag{40}$$

where $C_1$ is the correction factor employed to scale the heat transfer magnitude with the value calculated from CFD study at steady state. Thus, the complete transient variation of the heat transfer coefficient in the vertical CNCL is assumed to be given by equation-41.

$$h_{1/2,VerticalCNCL}(t) = max\big(h_{1/2,conduction}(t), h_{1/2,VerticalHX}(t)\big) \tag{41}$$

The motivation for this approach was taken from a study conducted by Hellums and Churchill [15] on the transient free convection heat transfer on an isothermal flat plate.

The heat transfer in the heat exchanger section of the horizontal CNCL system is determined by the flow initiated by the buoyancy forces generated in the vertical insulated legs of the component NCLs. Hence, the forced convection correlations are utilised to model the heat transfer coefficient in the heat exchanger section (horizontal leg). The heat transfer coefficient variation in horizontal heat exchanger is represented in the most general form by equation-42 and the transient heat transfer coefficient is given by equation-43.

$$h_{1/2, HorizontalHX}(t) = C_2 Re(t)^{1/2} \tag{42}$$

where $C_2$ is another correction factor which is used to scale the heat transfer value obtained from CFD steady state analysis.

$$h_{1/2, HorizontalCNCL}(t) = max\big(h_{1/2, conduction}(t), h_{1/2, HorizontalHX}(t)\big) \tag{43}$$

The overall heat transfer coefficient (U) at the heat exchanger section is evaluated using equation-44.

$$U = \frac{h_1 h_2}{h_1 + h_2} \tag{44}$$

The study conducted in section.4.4 on the effect of heater and cooler configurations on the CNCL system indicates that the magnitude of heat transfer coefficient $U$ for both the horizontal and vertical CNCL system, lies in the range: 18,000-29,500 $W/m^2 K$. The influence of variation of $U$ on the CNCL system is studied and represented by Fig.14 and it indicates that the system dynamics is not greatly influenced by the heat transfer coefficient.

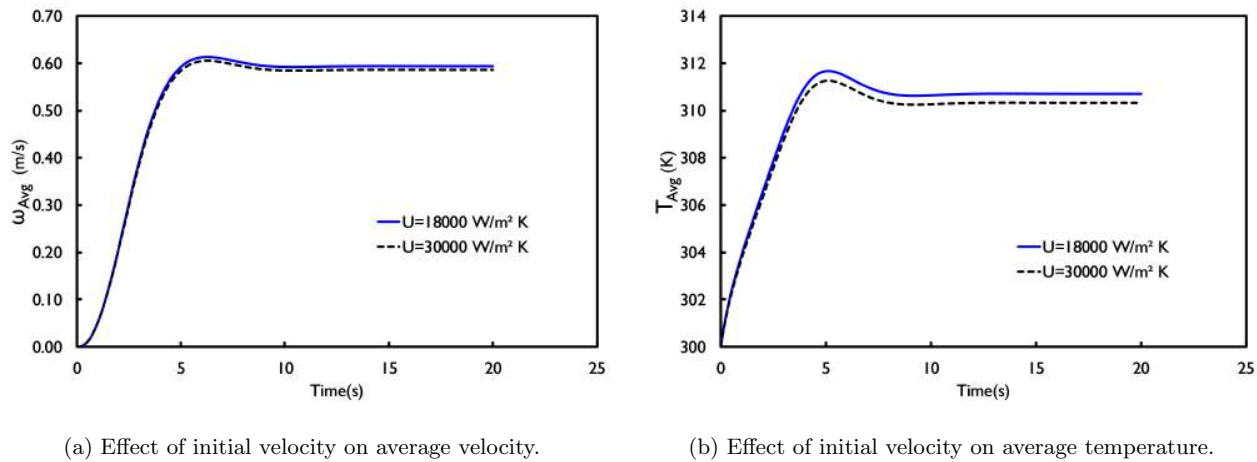

(a) Effect of initial velocity on average velocity.

(b) Effect of initial velocity on average temperature.

Figure 14: Effect of the heat transfer coefficient variation on CNCL-(a) using 1-D model with the critical parameters ($Q''$, $L$, $L1$, $D_h$) held constant (Table.3 and Table.5) .

For the fluids considered in the present study (FF1 and FF2), the thermophysical properties are closer to the magnitudes of liquid metals. For liquid metals, the heat transfer coefficient is determined by the relation:

$$Nu = A + B(RePr)^n \tag{45}$$

where, A, B, n are constants with $A \in (5-7)$, $B \in (0.01-0.03)$ and $n \in (0.4-0.8)$. Thus for liquid metals, the heat transfer coefficient is not much dependent on the fluid considered or other factors significantly as the value of constant 'A' determines the order of magnitude of the heat transfer coefficient. Mochizuki [16] performed a comparative test on the different heat transfer correlations available to predict the heat transfer at low Reynolds number. He concluded that the correlation stated by Seban and Shimazaki [17] is ideal for predicting heat transfer at low Reynolds number. The Nusselt number correlation proposed by Seban and Shimazaki for liquid sodium is

$$Nu = 5 + 0.025(RePr)^{0.8} \tag{46}$$

The above correlation has been developed for accounting the heat transfer across the entire surface area of the pipe engaged in heat exchange. However, the heat exchange region of the current study is a flat plate type heat exchanger, hence the scaled version of equation-46 is used which is represented by equation-47.

$$Nu = \frac{\pi D_h}{4L}\left(5 + 0.025(RePr)^{0.8}\right) \tag{47}$$

*5.2. Validation of the 1-D CNCL model*

*5.2.1. Without considering the bend loss effect in the 1-D CNCL model*

Incorporating the aforementioned correlations for friction factor and heat transfer coefficient, the transient dynamics of the CNCL system has been simulated using 1-D model and the results have been compared with 3-D CFD study to verify the 1-D model, as shown in Fig.15.

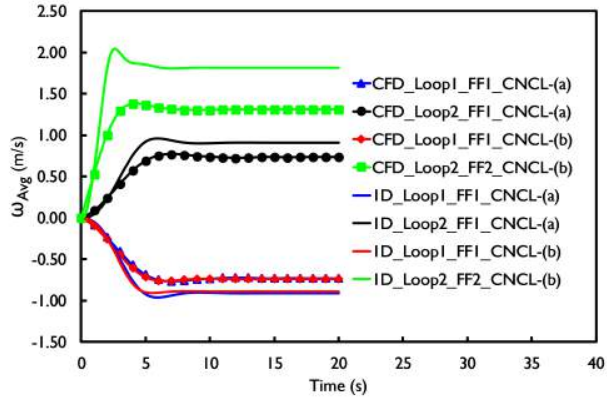
(a) Average velocity vs time for CNCL-(a) and CNCL-(b).

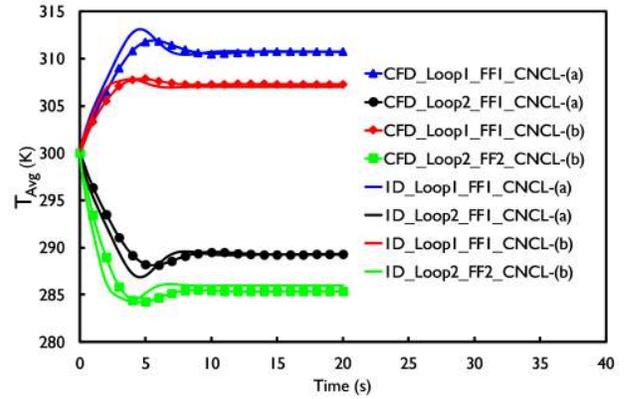
(b) Average temperature vs time for CNCL-(a) and CNCL-(b).

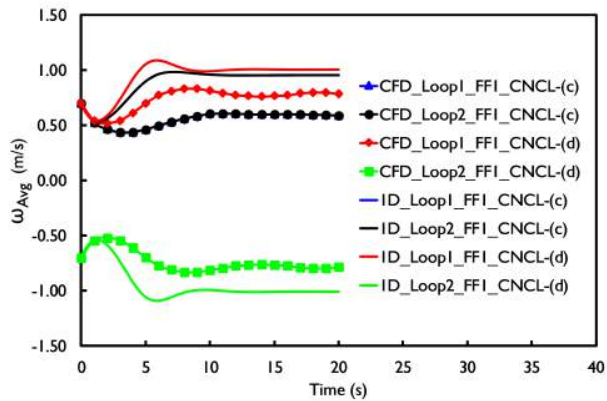
(c) Average velocity vs time for CNCL-(c) and CNCL-(d).

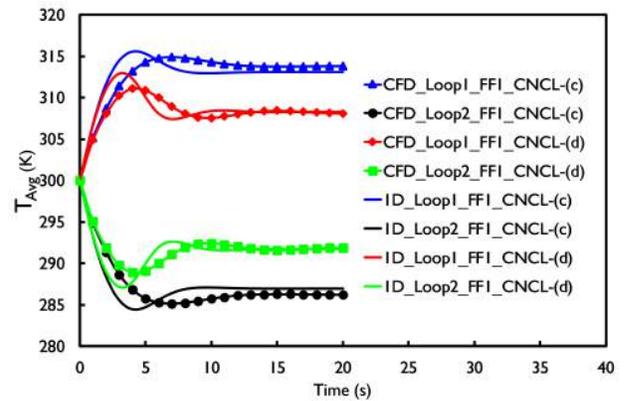
(d) Average temperature vs time for CNCL-(c) and CNCL-(d).

Figure 15: Comparison of the 1-D model vs the 3-D CFD model of the CNCL system without bend loss consideration. The lines with markers indicate CFD values and the plain line indicates 1-D prediction.

From Fig.15a and Fig.15c we observe that the actual trend of the velocity obtained from CFD is properly captured by the 1-D model even though it over predicts the velocity for all the cases considered. There is a good match between the CFD average velocity plots and those of the 1-D model during the initial transience, but this deteriorates with passage of time. Before explaining the cause of the deviation let us, look at the average temperature plots.

Figures 15b and 15d represent the average temperature variation of the fluid in loop 1 and 2 of the CNCL system, and there seems to be an excellent match between the CFD and 1-D results at the steady state and the actual trend of the plot is also properly captured by the 1-D model. The plot of average temperature of the 1-D model is slightly out of phase with the CFD predicted plot.

The average velocity plots obtained using the 1-D model clearly over-predict the velocity, but it is known that the energy supplied to a single phase NCL is stored in the form of its velocity (kinetic energy) and temperature (internal energy). From Fig.15b and Fig.15d it is observed that there is an excellent match between the CFD and 1-D results of the average temperature plots , thus the over-prediction in average velocity observed in Fig.15a and Fig.15c indicates that there is a drop in kinetic energy of the system predicted using CFD simulation. How the remaining kinetic energy is expended can be explained by having a closer look at the velocity contours, streamlines and velocity vectors at the bends which indicate that the kinetic energy is lost due the vortices(Fig.12), thus what we observe is a transformation of the translation kinetic energy of the system to rotational energy of the vortices and eddies. The 1-D model does not take into account the flow losses at the bends, thus causing an apparent over-prediction of the 1-D model for average velocity plots.

From Fig.15a and Fig.15c we find that the maximum over-prediction in average velocity plot corresponds to the parallel flow heat exchanger. This can be attributed to the non-uniform temperature drop across the heat exchanger section in the parallel flow heat exchanger. Fluids that are used in industrial and everyday applications have low velocity magnitudes when activated solely by buoyancy forces as in the case of the NCL or CNCL system.

5.2.2. Incorporating bend loss effects into the 1-D CNCL model

For liquids which are used in day to day applications the bend losses may not play a role on the transient behaviour of a buoyancy driven system, but for fluids with high thermal conductivity such as liquid metals and the fictitious fluids considered for the present study, the bend loss is an important parameter as discussed in section 4.3. Tarantino et al [18] and Borgohain et al. [19] employed the bend loss factors in their study pertaining to liquid metals. To model the bend losses we incorporate an additional pressure loss $\Delta P_{bend}$ which is calculated using equation-49; we shall also employ the assumption of a bend loss coefficient independent of Reynolds number ([20]).

The momentum equation after incorporating the bend loss is given as:

$$\rho_{1,2}\frac{d\omega_{1,2}(t)}{dt} + \frac{4(\tau_{1,2} + \Delta P_{bend})}{D_h} = \frac{\rho_{1,2}g\beta_{1,2}}{2(L+L1)}(\oint(T_{1,2}-T_0)f_{1,2}(x)dx) \qquad (48)$$

where $\Delta P_{bend}$ is the pressure loss at the 90 degree elbow bends.

$$(\Delta P_{bend})_{1,2} = \frac{1}{2}K\rho_{1,2}\omega_{1,2}^2 \qquad (49)$$

and $K$ is the bend loss coefficient.

When the bend loss coefficient for FF1 was set to 0.15 and 0.4 for FF2, the agreement between the CFD and 1-D results is best. Thus we observe an excellent agreement between the CFD and 1-D model is obtained after incorporating the bend loss effects into the 1-D model of the CNCL system as observed in Fig.16. This indicates the importance of including the bend loss effects when studying the buoyancy driven flows in liquid metals, which corresponds with the literature surveyed and listed in Table.9.

The literature tabulated in Table.9 also indicates the widespread use of 1-D methods to understand buoyancy driven flows during transience and at steady state, thus strengthening the importance of the current study and its

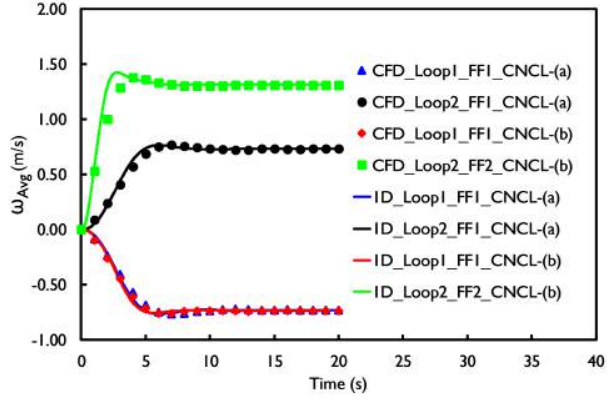
(a) Average velocity vs time for CNCL-(a) and CNCL-(b).

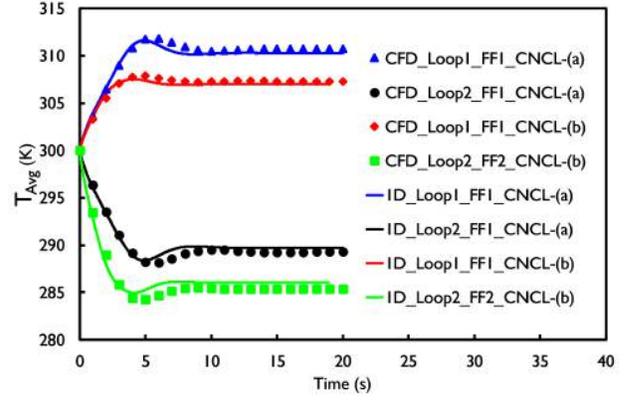
(b) Average temperature vs time for CNCL-(a) and CNCL-(b).

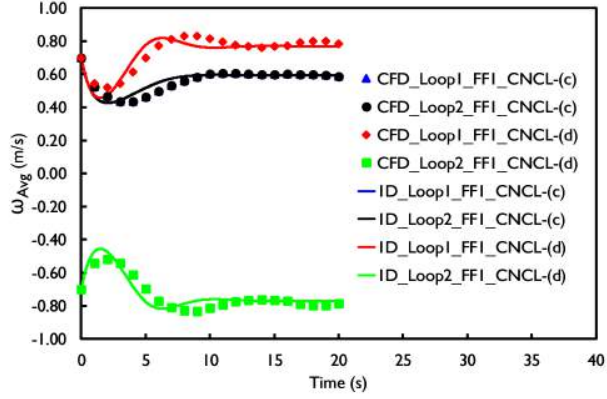
(c) Average velocity vs time for CNCL(c) and CNCL-(d).

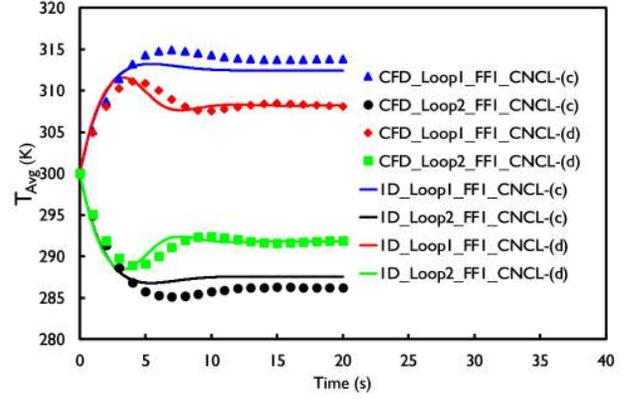
(d) Average temperature vs time for CNCL-(a) and CNCL-(b).

Figure 16: Comparison of the 1-D model vs the 3-D CFD model of the CNCL system after incorporating the bend losses. The markers indicate CFD results and the plain line represents the 1-D model prediction.

Table 9: Literature on liquid metal natural circulation with bend loss effects considered in the study.

| SI.No | Reference | Type of analysis | Method of study(bend loss incorporated) |
|---|---|---|---|
| 1 | Shin et al. (2015) [21] | Steady state | 1-D computational analysis |
| 2 | Borgohain et al. (2016) [22] | Transient | 1-D finite difference method |
| 3 | Son and Suh (2014) [23] | Transient | 1-D finite difference method |
| 4 | Takahashi et al. (2005) [24] | Transient | 1-D analytical |
| 5 | Srivastava et al. (2016) [25] | Transient | In-house code (LeBENC) |
| 6 | Shin et al. (2017) [26] | Steady state | MARS safety analysis code |
| 7 | Tarantino et al. (2008) [18] | Steady state | 1-D analytical |

practical relevance for industrial and engineering applications.

### 5.3. Validation of the modelling methodology

This section deals with the verification of the 1-D modelling methodology employed to model the CNCL system. From Fig.3, it is evident that the modelling of Loop 1 of the CNCL system is similar to the modelling of a single NCL with heat flux input at the heater section and a transient temperature condition at the cooler section.

Thus to verify the modelling methodology, we can use the current 1-D modelling approach to model an NCL with a heat flux boundary condition at the heater section but a constant temperature condition at the cooler section. The reason for selection of the NCL ($+Q''$ and constant Temperature boundary condition) for validating the modelling methodology is the availability of literature on such systems.

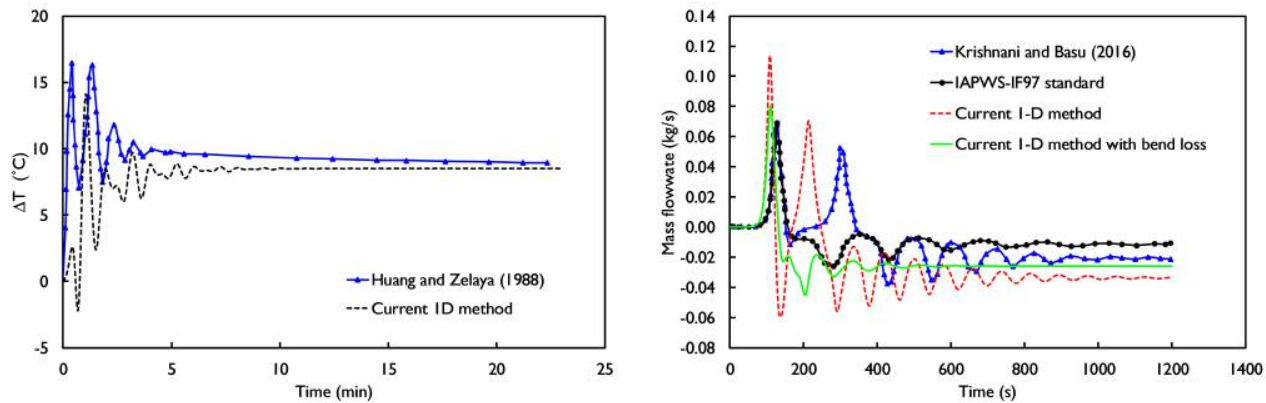

(a) Comparison of the current 1-D method with 1-D finite difference model of NCL with water as the operating fluid for 1200 $W$.

(b) Comparison of the current 1-D method with 3-D CFD study with water as the operating fluid for 500 $W$ and constant temperature of 275 $K$.

Figure 17: Validation of the modelling methodology used to model the CNCL system.

Figures 17a and 17b represent the comparison of the CNCL modelling methodology with 1-D finite difference model by Huang and Zelaya [27] and 3-D CFD study undertaken by Krishnani and Basu [28], respectively.

Figure 17a shows a good agreement of the current methodology with the 1-D finite difference model used by Huang and Zelaya. The heater and cooler sections of the NCL considered are on the bottom and top parts of the vertical legs of the NCL respectively. $\Delta T$ represents the difference between the inlet and exit temperatures of the operating fluid flowing through the cooler section. The deviation observed in the transient behaviour can be attributed to minor differences in the thermophysical properties used for the validation purpose.

Figure 17b represents the comparison of current methodology with 3-D CFD study. Krishnani and Basu employed Boussinesq approximation for the 3-D CFD study and compared their results with those obtained using the IAPWS-IF97 standards [29]. The NCL considered has the heater and cooler sections on the horizontal legs.

The following observations can be made from Figures 17a and 17b:

1. It has been observed that the magnitude of the overall heat transfer coefficient ($U$) and the bend loss coefficient ($K$) determine the morphology of the transient behaviour of NCL system. This can explain the differences observed in Figures 17a and 17b.

2. The morphology of the transient plots employing the current methodology with bend loss are in good agreement with 3-D CFD study conducted by Krishnani and Basu employing the IAPWS-IF97 standards.

Thus, the modelling methodology employed in the present study has been verified using the results obtained in literature (1-D finite difference and 3-D CFD) and it can be concluded that the CNCL modelling is reliable. However the magnitudes of the thermophysical properties along with the magnitude of $U$ and $K$ determine the transient morphology of the system studied. This conclusion is especially valid for systems exhibiting oscillatory behaviour.

## 6. Parametric study of the CNCL system employing the 1-D model

This section provides a detailed parametric study on the effect of initial conditions, hydraulic diameter ($D_h$), input heat flux ($Q''$) and aspect ratio ($L/L1$) of the CNCL system. The CNCL system considered for the study has its constituent loops filled with the same fluid and the magnitude of average fluid velocity and the average temperature of the heater loop are used to characterize the CNCL system behaviour. The 'h1c1' configuration for vertical CNCL system and the counterflow 'h2c2' configuration for the horizontal CNCL system is utilised for carrying out the parametric studies with $K = 0$.

### 6.1. Effect of initial velocity on the CNCL system behaviour

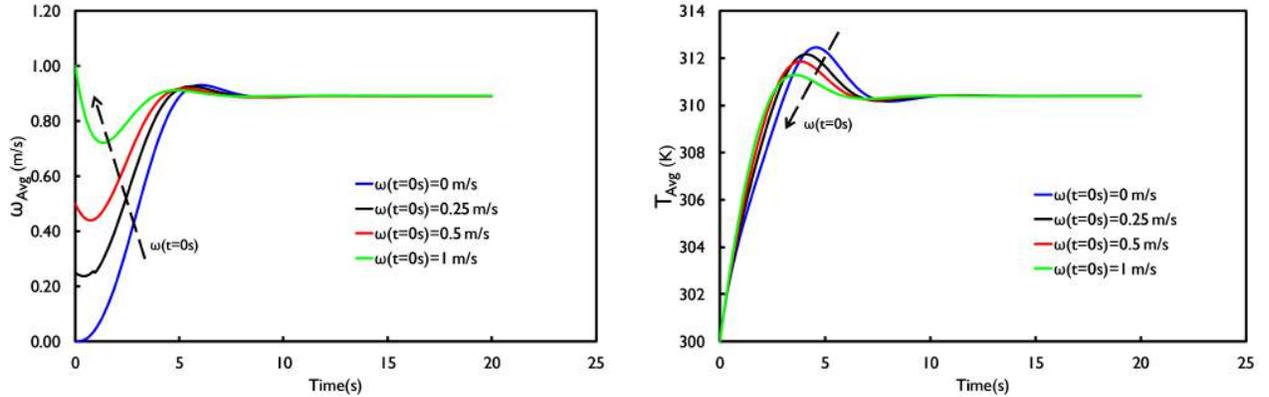

(a) Effect of initial velocity on average velocity of the CNCL system.

(b) Effect of initial velocity on average temperature of the CNCL system.

Figure 18: The effect of initial velocity on the behaviour of the vertical CNCL with both the loops containing FF1 for $D_h = 40\ mm$, $L = L1 = 1\ m$, $Q'' = 2000\ W/m^2$ and initial temperature of 300 K.

From Fig.18 we observe that for a vertical CNCL system with increase in the initial velocity the initial transient behaviour of the average velocity plot is significantly affected, whereas the time taken to attain steady state is hardly influenced. The average temperature plots indicate a gradual offset to the left along with the dampening of the mound of the transient curves with increase in the initial velocity.

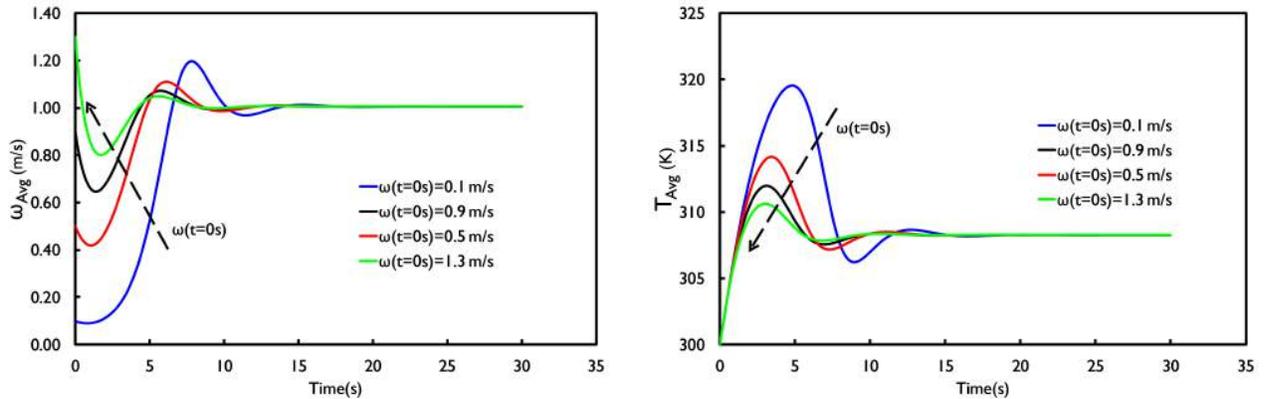

(a) Effect of initial velocity on average velocity of the CNCL system.

(b) Effect of initial velocity on average temperature of the CNCL system.

Figure 19: The effect of initial velocity on the behaviour of the horizontal CNCL with both the loops containing FF1 for $D_h = 40\ mm$, $L = L1 = 1\ m$, $Q'' = 2000\ W/m^2$ and initial temperature of 300 K.

Fig. 19 indicates that the initial velocity considerably affects the transient behaviour of the horizontal CNCL system. The initial transience of the average fluid velocity and temperature curves are substantially affected as the initial velocity is raised to non-zero velocity magnitudes along with the time taken to attain steady state. This indicates that providing an initial velocity to the horizontal system considerably reduces the time required to reach steady state. The offset and dampening of the transient average temperature plot is clearly visible, signifying the similarity in the overall trend of the horizontal and vertical CNCL system behaviour with increase in the initial velocity.

## 6.2. Effect of initial temperature on the CNCL system behaviour

Figures 20 and 21 represent the effect of change in the initial temperature on the CNCL system. It turns out that the initial temperature has no effect on the average velocity plot and leads to an upward shift in the average temperature profile for both the horizontal and vertical CNCL systems. Real fluids with temperature dependent thermophysical properties do exhibit a dependence on the initial temperature of the CNCL system, but it is beyond the scope of the current study.

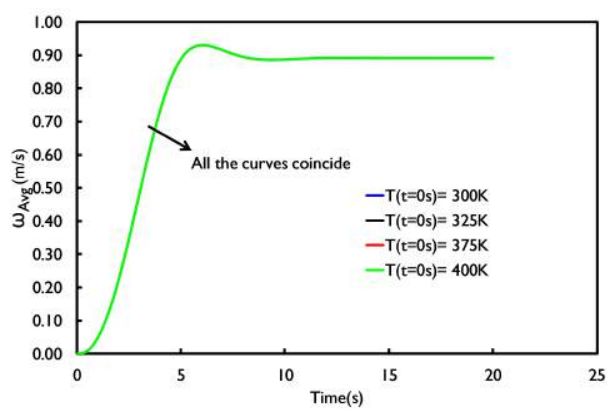
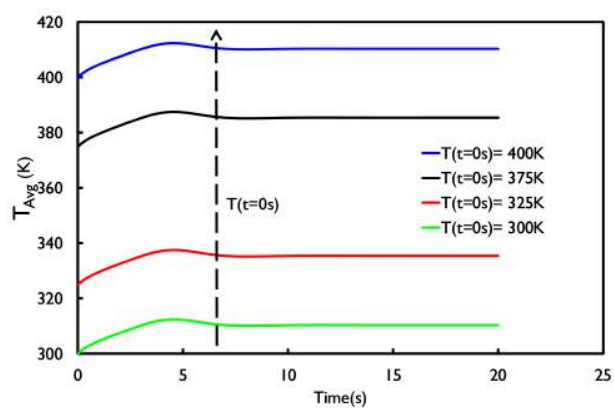

(a) Effect of initial temperature on the average velocity of the CNCL system.

(b) Effect of initial temperature on the average temperature of the CNCL system.

Figure 20: The effect of initial temperature on the behaviour of the vertical CNCL with both the loops containing FF1 for $D_h = 40$ mm, $L = L1 = 1$ m, $Q'' = 2000$ $W/m^2$ and initial velocity of 0 $m/s$.

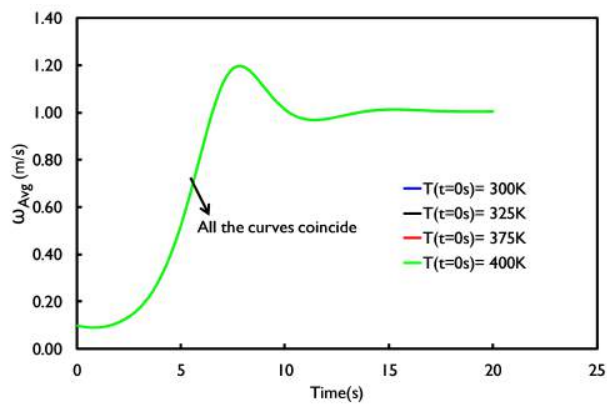
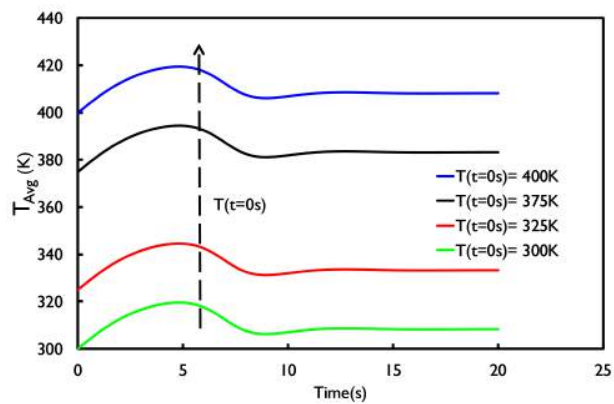

(a) Effect of initial temperature on the average velocity of the CNCL system.

(b) Effect of initial temperature on the average temperature of the CNCL system.

Figure 21: The effect of initial temperature on the behaviour of the horizontal CNCL with both the loops containing FF1 for $D_h = 40$ mm, $L = L1 = 1$ m, $Q'' = 2000$ $W/m^2$ and initial velocity of 0.1 $m/s$.

The reason for the peculiar characteristic of non-dependence on initial temperature is due to the fact that $\oint f(x) = 0$ for the considered geometry, where $f(x)$ is the piece-wise function which identifies the vertical and horizontal limbs of the CNCL w.r.t. gravity. This implies that if one of the legs of the constituent NCLs of the CNCL system were inclined then $\oint f(x) \neq 0$ then the dependence of the CNCL system on initial temperature could have been distinctly observed.

*6.3. Estimation of the time taken by the CNCL system to attain steady state*

It is important to be able to estimate the time a system requires to attain steady state as it assists in the design and maintenance of the system. However, from the aforementioned sections it is clear that the time taken to reach the steady state is dependent on its initial velocity, thus an approximate relation is proposed to find the time taken by system to reach steady state.

$$t_{ss} = \frac{2(L + L1)}{min(\omega_{1,ss}, \omega_{2,ss})} \tag{50}$$

Equation-50 only holds well when the CNCL system is initiated with zero initial velocity and gives a very conservative estimate of the time during which the system is in transient phase. So for all cases $t_{actual,ss} > t_{ss}$, where $t_{actual,ss}$ is the actual amount of time the system requires to attain steady state and $t_{ss}$ is predicted using equation-50.

*6.4. Effect of variation of hydraulic diameter on the CNCL system behaviour*

From this section onwards, we shall consider the CNCL with operating fluid as sodium (Na) and use equation-47 to evaluate the heat transfer coefficient at the heat exchanger section. This is done as the rest of the parametric studies conducted modify the dimensions and boundary conditions of the CNCL system, thus the prediction the transient dynamics is possible only if the heat transfer coefficient (which depends on the mentioned parameters) can be estimated at the heat exchanger section. Table.10 indicates the thermophysical properties of liquid Sodium (Na) at $200°C$ which are used for all the upcoming parametric studies.

Table 10: Thermophysical properties of liquid soduim(Na) at $473.15K$ .

| Property | Liquid Sodium(Na) |
|---|---|
| $\rho_0$ | $902.809 \ kg/m^3$ |
| $C_p$ | $1343.123 \ J/kgK$ |
| $\beta$ | $2.602 \times 10^{-4} 1/K$ |
| $a$ | $6.757 \times 10^{-5} m^2/s$ |
| $\mu$ | $4.46 \times 10^{-4} kg/ms$ |

As will be observed from the upcoming section (section 6.7) on the effect of heat flux on the CNCL system behaviour with Sodium as the working fluid that with increasing heat flux the oscillatory behaviour of the system dominates, eventually leading to chaotic aperiodic oscillations. Thus, to obtain a proper understanding of the coupling between velocity and temperature, low heat flux values are utilised to conduct the parametric studies.

Figures 22 and 23 represent the effect of hydraulic diameter on the CNCL system. Both the vertical and horizontal system indicate similar trends for the corresponding transient average velocity and temperature plots. The increase in the transient average velocity plot with hydraulic diameter is due to the increase in the area of the heating and cooling section for the same heat flux magnitude, whereas the decline in the average temperature plot can be attributed to the steep addition of mass (heat storing capacity) to the CNCL compared to heat input.

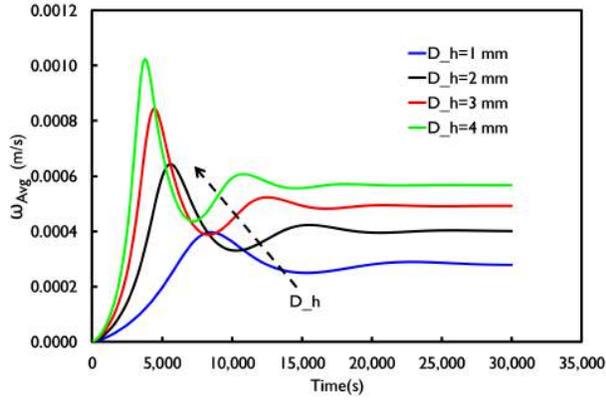
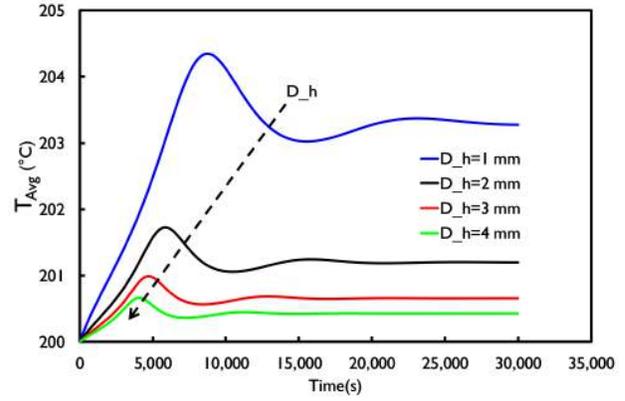

(a) Effect of hydraulic diameter on the average velocity.

(b) Effect of hydraulic diameter on average temperature.

Figure 22: Effect of the hydraulic diameter on the average fluid velocity and temperature for liquid sodium filled loops at 473.15 K with $L = L1 = 1$ m, $Q'' = 1\ W/m^2$ on the vertical CNCL system.

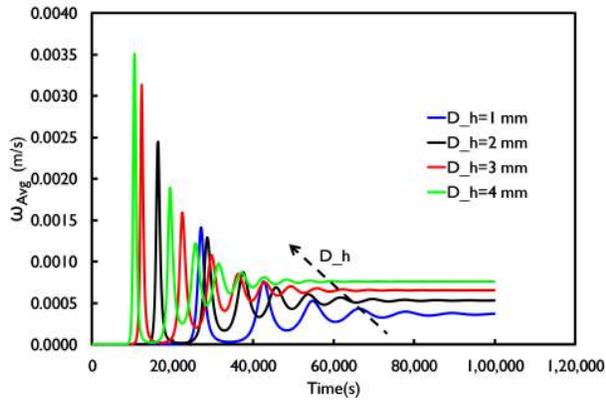
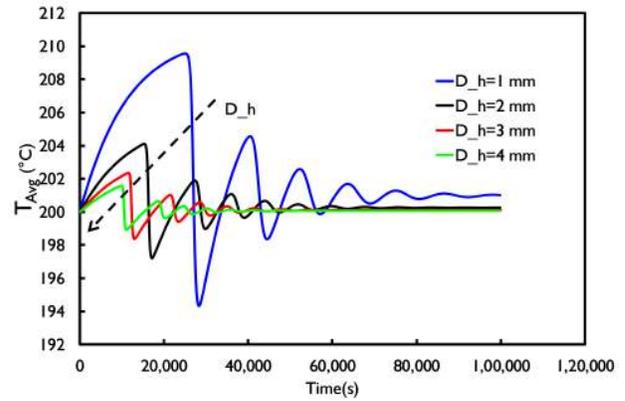

(a) Effect of hydraulic diameter on average velocity.

(b) Effect of hydraulic diameter on average temperature.

Figure 23: Effect of the hydraulic diameter on the average fluid velocity and temperature for liquid sodium filled loops at 473.15 K with $L = L1 = 1$ m, $Q'' = 1\ W/m^2$ on the horizontal CNCL system.

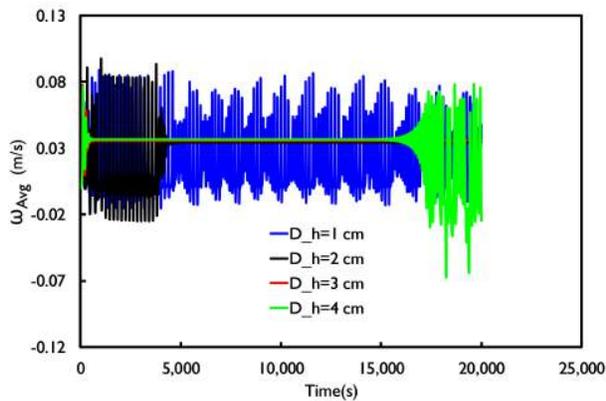
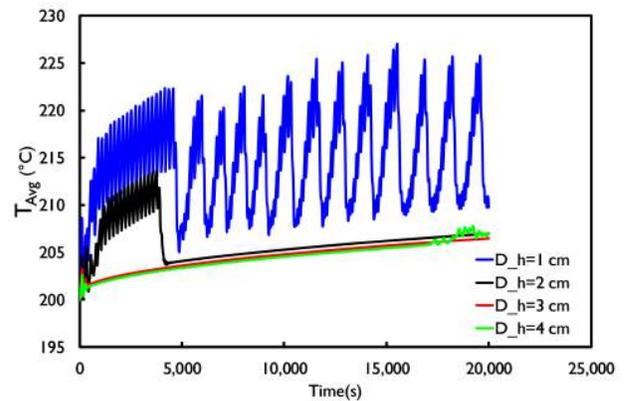

(a) Effect of hydraulic diameter on average velocity.

(b) Effect of hydraulic diameter on average temperature.

Figure 24: Effect of the hydraulic diameter on the average fluid velocity and temperature for liquid sodium filled loops at 473.15 K with $L = L1 = 1$ m, $Q'' = 1000\ W/m^2$ on the horizontal CNCL system.

Fig.24 represents the effect of varying the hydraulic diameter at high heat loads. It is observed that the system exhibits significant oscillatory behaviour at high heat loads and no observable trend can be deciphered from Fig.24a.

Figure24b representing average temperature transient behaviour at high heat loads shows a similar trend as Fig.22b.

Similar trends as observed at low power loads are exhibited by both the vertical and horizontal system at higher power loads too (wherever the trend is discernible). Thus for the sake of brevity only the CNCL behaviour at low power load is presented.

*6.5. Effect of variation of heat flux on the CNCL system behaviour*

Figures 25 and 26 reveal that both the average velocity and temperature increase with time with the escalation of heat flux input for the vertical and horizontal CNCL systems. This behaviour can be directly ascribed to the increase in the buoyancy forces with rise in the heat flux, as the other parameters are held constant.

The system behaviour at higher heat flux magnitudes exhibits uni-directional or bi-directional flow reversals with chaotic behaviour that is elaborately discussed in the *Section* 6.7.

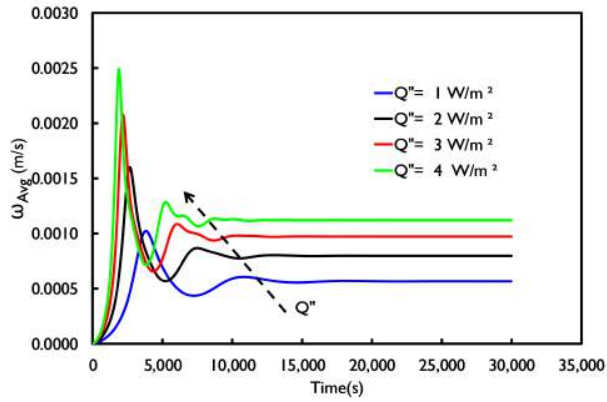
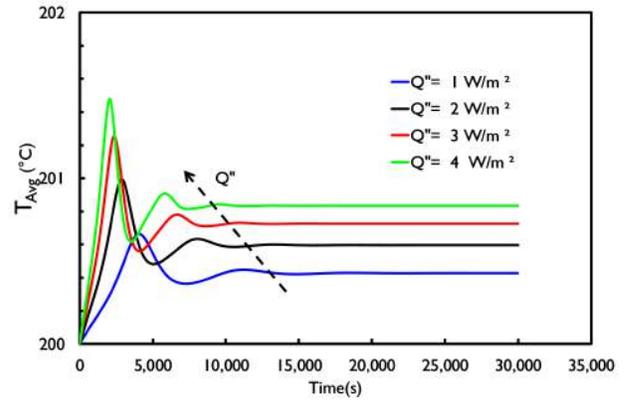

(a) Effect of heat flux on the average velocity of the CNCL system.

(b) Effect of heat flux on the average temperature of the CNCL system.

Figure 25: Effect of the heat flux on the average fluid velocity and temperature for liquid sodium filled loops at 473.15 K with $L = L1 = 1$ m, $D_h = 4$ mm on the vertical CNCL system.

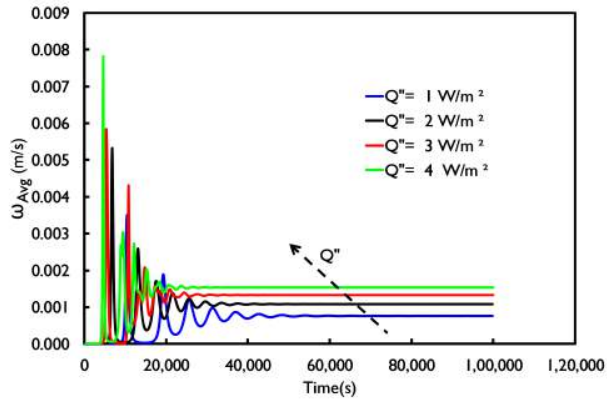
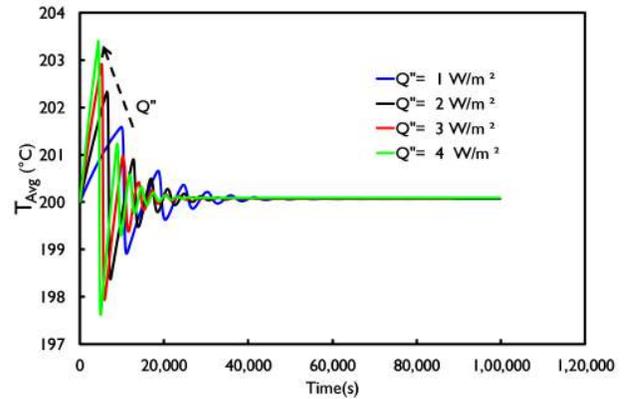

(a) Effect of heat flux on the average velocity of the CNCL system.

(b) Effect of heat flux on the average temperature of the CNCL system.

Figure 26: Effect of the heat flux on the average fluid velocity and temperature for liquid sodium filled loops at 473.15 K with $L = L1 = 1$ m, $D_h = 4$ mm on the horizontal CNCL system.

## 6.6. Effect of variation of aspect ratio on the CNCL system behaviour

The aspect ratio of the CNCL is defined as the ratio $L/L1$. The aspect ratio determines the geometry of the constituent loops of the CNCL system. For performing the current parametric study on the effect of aspect ratio on CNCL system behaviour we have fixed the value of $L1$ as $1\ m$, thus the aspect ratio is varied solely by modifying the height of the CNCL ($L$).

From Fig.27 and Fig.28 we observe that the transient average velocity plots of the horizontal and vertical CNCL system ascend with increasing aspect ratio due to increase in the heating and cooling section area with increase in height of the CNCL system. This leads to an increase in heat flux and the buoyancy forces, which go hand in hand.

We observe that increasing the aspect ratio leads to increase in the heat input to the CNCL system, but at the same time causes a relatively larger increase (w.r.t. heat input) in the mass (heat storing capacity) of the system which leads to a drop in the average temperature of the vertical CNCL system. The average temperature plot of the horizontal CNCL system does not indicate much change in the steady state behaviour because of variation of aspect ratio, but the peak of the plots increase with ascending aspect ratio. This maybe because of the initial unequal trade-off between the increase in the heat input rate and heat storing capacity rate during the initial transience.

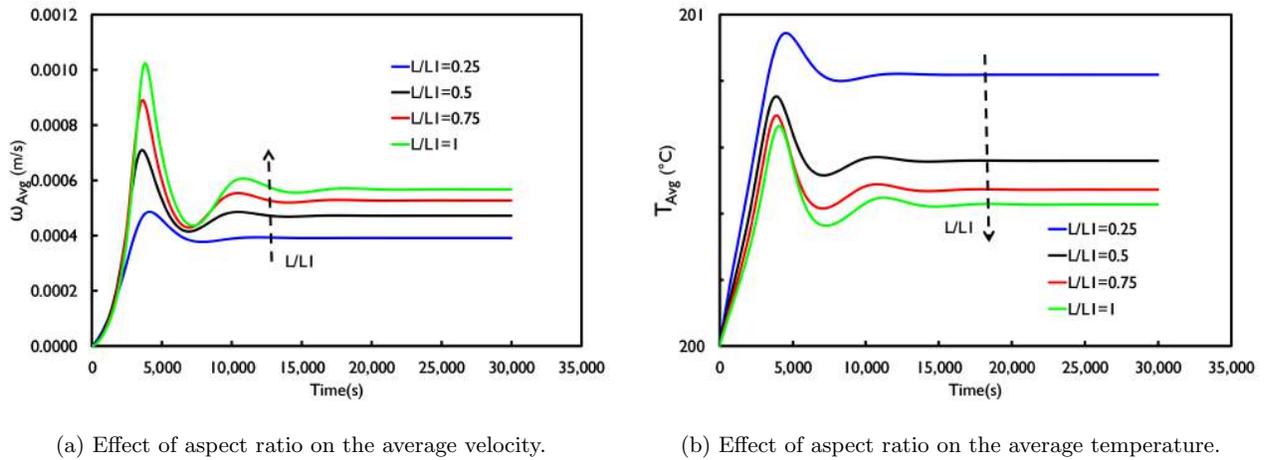

(a) Effect of aspect ratio on the average velocity.

(b) Effect of aspect ratio on the average temperature.

Figure 27: Effect of the aspect ratio on the average fluid velocity and temperature for liquid sodium filled loops at $473.15\ K$ with $Q'' = 1\ W/m^2$, $D_h = 4\ mm$ on the vertical CNCL system.

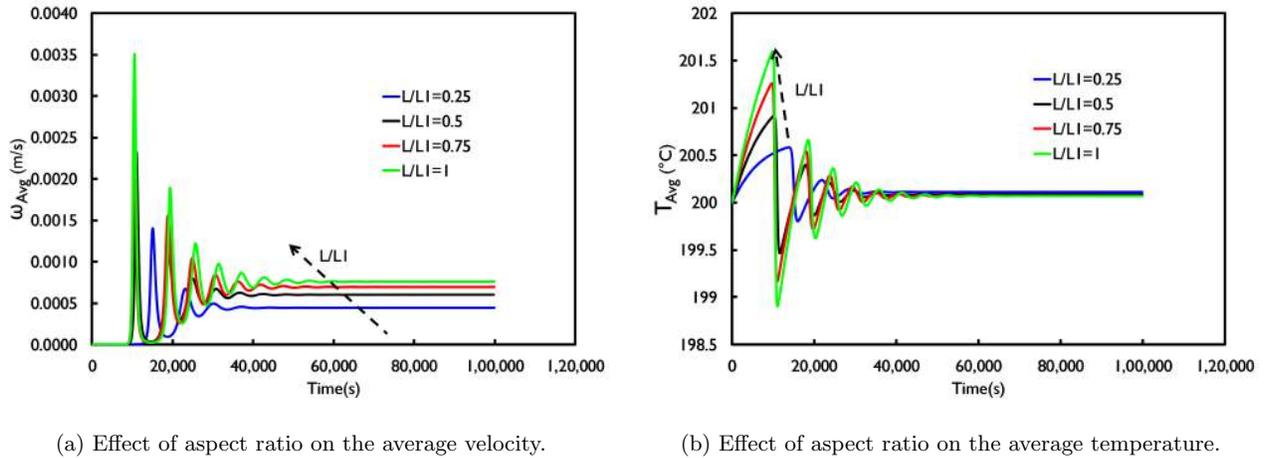

(a) Effect of aspect ratio on the average velocity.

(b) Effect of aspect ratio on the average temperature.

Figure 28: Effect of the aspect ratio on the average fluid velocity and temperature for liquid sodium filled loops at $473.15\ K$ with $Q'' = 1\ W/m^2$, $D_h = 4\ mm$ on the horizontal CNCL system.

*6.7. Flow characteristics of CNCL systems at higher heat loads*

This section is an extension of the section 6.5 (effect of heat flux). At low heat input, the system attains a steady state with or without initial oscillatory transient behaviour that dies out gradually, but at high heat loads, the oscillations fail to die down leading to uni or bi-directional oscillations along with chaotic or complete non-periodic behaviour in some cases as represented by Fig.30. This property of the CNCL system can be ascribed to the behaviour of its component NCLs at high heat loads, this behaviour of the NCLs at high heat loads was explained by Welander [4] and was attributed to the buoyancy and viscous forces not being in phase, which result in oscillatory or chaotic oscillations.

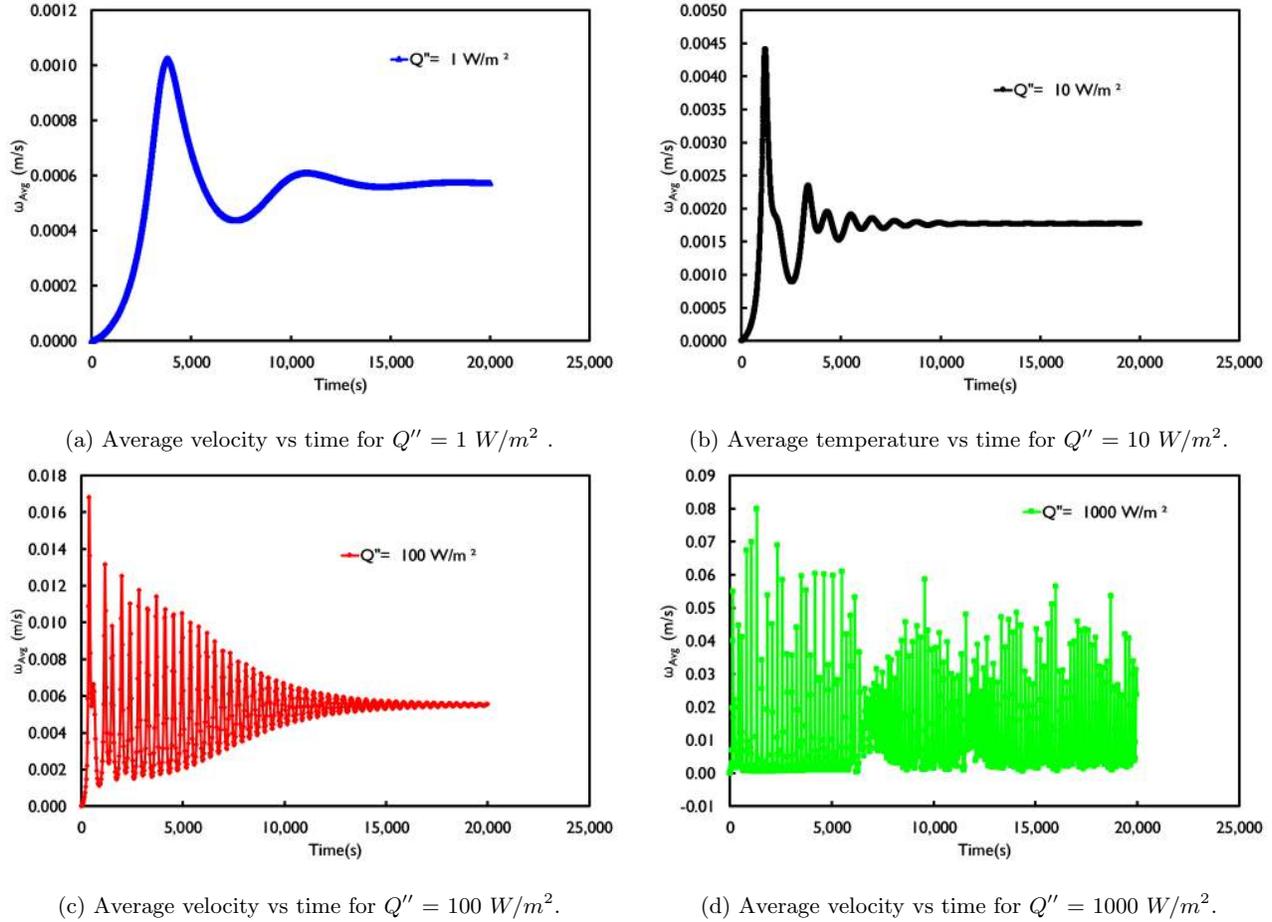

(a) Average velocity vs time for $Q'' = 1\ W/m^2$ .

(b) Average temperature vs time for $Q'' = 10\ W/m^2$.

(c) Average velocity vs time for $Q'' = 100\ W/m^2$.

(d) Average velocity vs time for $Q'' = 1000\ W/m^2$.

Figure 29: Effect of different orders of magnitudes of heat flux on the vertical CNCL system with $D_h = 4\ mm$, $L = L1 = 1$ m .

Figure 29 shows the effect of increasing heat flux by orders of magnitude on the CNCL system. With increasing heat flux, the oscillatory nature of the system manifests and eventually leads to aperiodic oscillatory behaviour. The average fluid flow velocity also increases with progressive increase in the heat load.

Figure 29d portrays the uni-directional (net flow in one direction) oscillations induced in the system for the given geometric dimensions and the fluid used for the analysis. The average flow is still either in clockwise or anti-clockwise direction. Fig.30 displays the bi-directional oscillations observed in the vertical CNCL system where the average flow is nearly zero and only oscillations dominate. Similar behaviour is also exhibited by simple NCL systems too and was observed experimentally and theoretically (using the 1-D model of the NCL) by Fichera and Pagano [3].

Figure 31 shows the effect of increasing the heat flux by orders of magnitude on the horizontal CNCL system. It is observed that with increasing heat flux the system attains steady state much more rapidly relative to the vertical CNCL system along with increase in the magnitude of the steady state velocity.

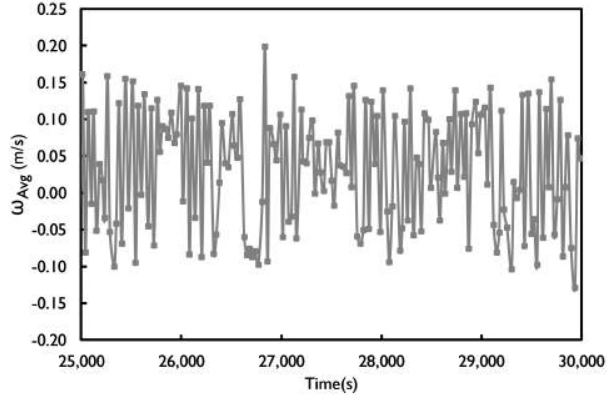
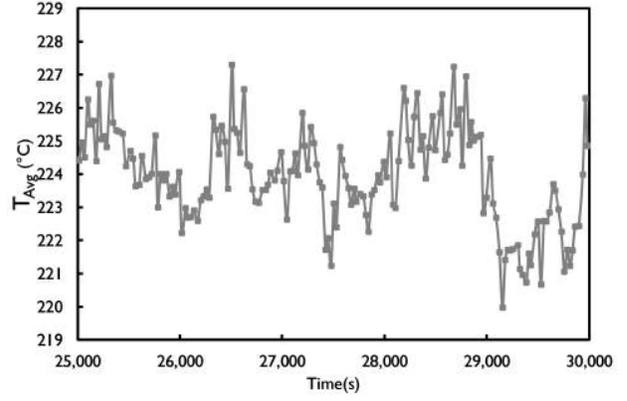

(a) Average velocity at high heat loads

(b) Average velocity at high heat loads

Figure 30: Effect of high heat load on the average fluid velocity and temperature for liquid sodium filled loops at $473.15\ K$ with $Q'' = 10000\ W/m^2$, $D_h = 40\ mm$, $L = L1 = 1$ m of the vertical CNCL system.

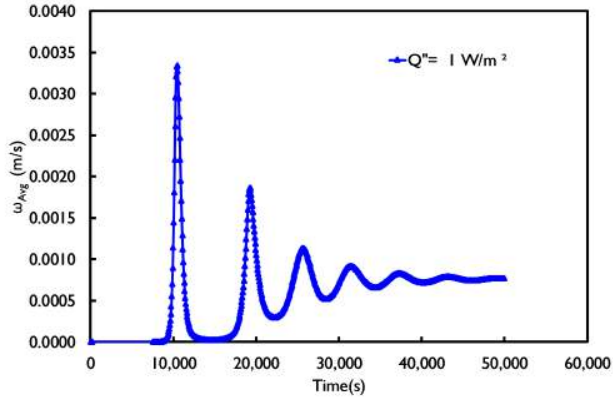
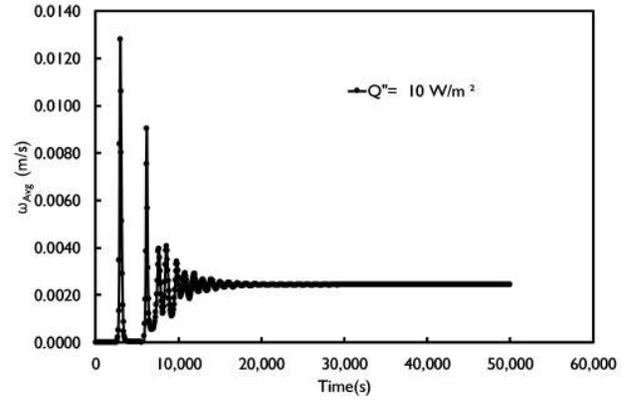

(a) Average velocity vs time for $Q'' = 1\ W/m^2$.

(b) Average velocity vs time for $Q'' = 10\ W/m^2$.

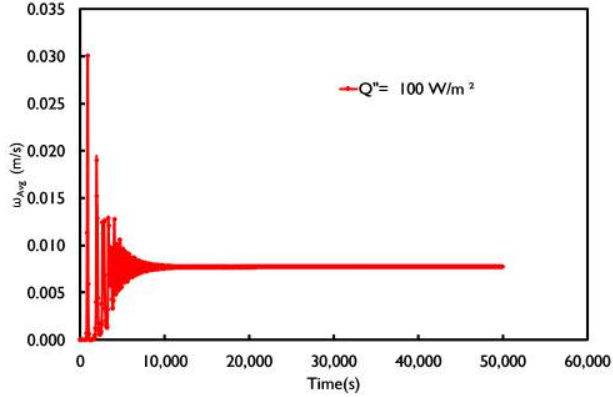
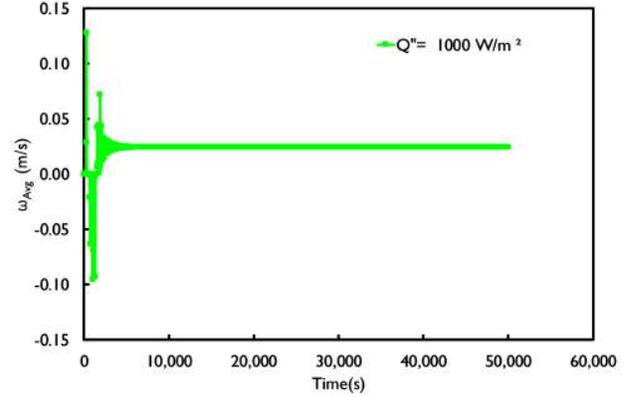

(c) Average velocity vs time for $Q'' = 100\ W/m^2$.

(d) Average velocity vs time for $Q'' = 1000\ W/m^2$.

Figure 31: Effect of different orders of magnitudes of heat flux on the horizontal CNCL system with $D_h = 4\ mm$, $L = L1 = 1$ m .

Comparing the figures 29 and 31 we find that the horizontal CNCL system tends to be more stable than the vertical CNCL system at high heat loads for the considered geometry. Thus to provide a definitive answer of stating whether a horizontal or vertical system is more stable a thorough stability analysis needs to conducted. The stability analysis of the CNCL system is beyond the scope of the current work and will be explored comprehensively in the future studies.

## 6.8. Comparison of an NCL system with a CNCL system

This section presents the similarities and differences amongst a single phase NCL and CNCL system. FF1 and Sodium are used as the operating fluids with constant heat flux boundary conditions for the heater and cooler configurations represented in Fig.32. The 1-D models of the NCL ([3]) and CNCL (present study) are used for illustrating the properties of the respective systems with the initial velocity of 0.1 $m/s$ and initial temperature of 300 $K$ for FF1 and 473.15 $K$ for Sodium.

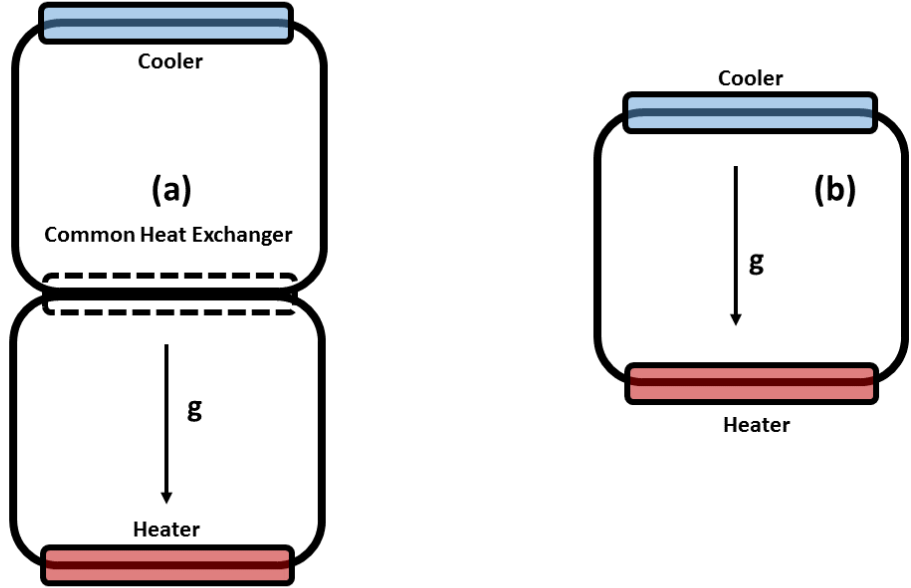

Figure 32: (a) Horizontal CNCL system (b) NCL system.

From Fig.33 we observe that the NCL system has the largest magnitude of average velocity relative to the horizontal CNCL with parallel-flow and counter-flow configuration at the heat exchanger section. The average temperature plot of NCL system does not vary with time unlike the horizontal CNCL. The coupling between the average temperature and velocity of the horizontal CNCL system evident from Fig.33.

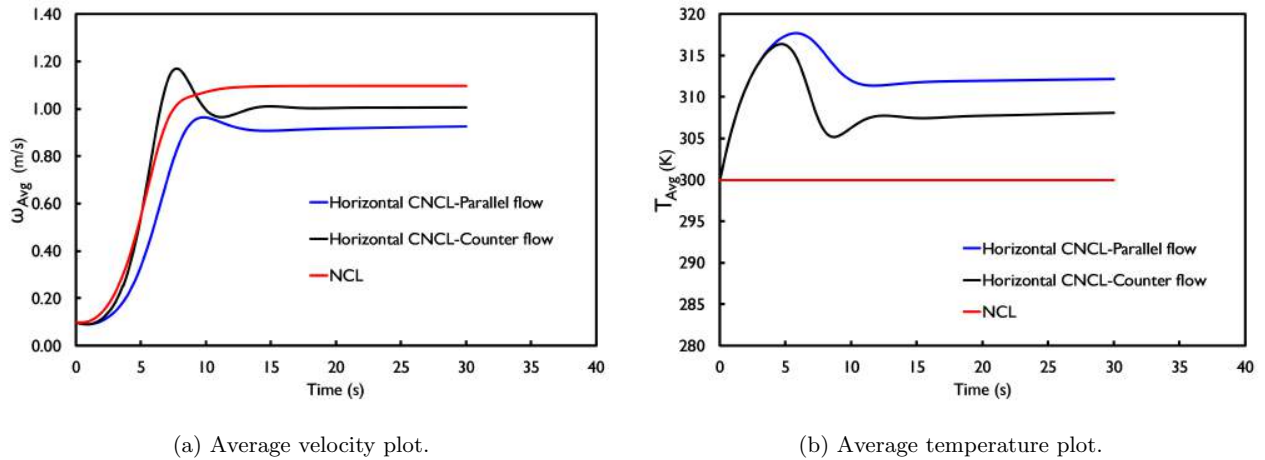

(a) Average velocity plot.

(b) Average temperature plot.

Figure 33: Comparison of the NCL system with horizontal CNCL system for $Q'' = 2000 \ W/m^2$, $D_h = 40 \ mm$, $K = 0$ and $L = L1 = 1 \ m$ with FF1 as the operating fluid.

Figure 34 represents the comparison of NCL with CNCL for Sodium as the operating fluid. It is observed that the NCL breaks into chaotic oscillation quickly relative to the horizontal CNCL system for the given heat load.

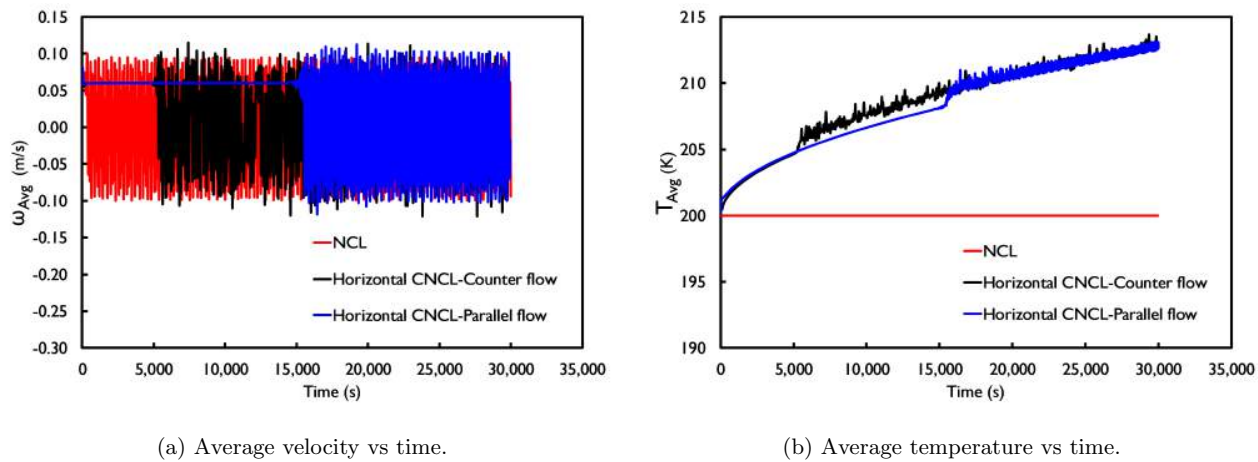

(a) Average velocity vs time.

(b) Average temperature vs time.

Figure 34: Comparison of the NCL system with horizontal CNCL system for $Q'' = 2000\ W/m^2$, $D_h = 40\ mm$, $K = 0$ and $L = L1 = 1\ m$ with Sodium as the operating fluid.

*6.9. Advantages and limitations of the 1-D single phase transient CNCL model*

Advantages of the 1-D CNCL model are:

1. Provides a simple framework to comprehend the complex behaviour exhibited by the CNCL system.
2. No commercial software is required to perform the numerical integration of the ODEs. (The current work uses MATLAB, but the same task can be accomplished by open-source software or through in-house code.)
3. Minimal computational power and time required to simulate the transient dynamics of the system.
4. All possible thermal boundary conditions (heat flux, temperature and volumetric heat generation) can be easily implemented.
5. Complicated geometries can be simulated subject to suitable assumptions and error margins.
6. An ideal tool for the design and parametric study of the CNCL system.

Some of the drawbacks of the 1-D model of the CNCL are:

1. The 1-D model is not self-sufficient to predict the entire transient behaviour of the CNCL system for common fluids. It requires the correction factors $C_1$ and $C_2$ from the CFD steady state study to determine the heat transfer coefficient magnitude at steady state.
2. The wall conduction is not considered in the 1-D model.

**7. Conclusions**

The transient study of a single-phase CNCL system has been conducted employing 1-D modelling and a 3-D CFD study. The transient study of the CNCL system is important, as it enables us to design heat exchangers, which employ buoyancy forces to fulfil the required heat transfer duty. As mentioned in the introduction section, a LMFBR and SDHW system can be considered as slightly complex models of the CNCL system. The CNCL system allows us to estimate the flow characteristics and temperature variations of an LMFBR system in off duty conditions, or simulate the system capability to handle the heat load during emergencies such as loss of coolant accidents.

Thus far the literature available on transient single-phase CNCL systems have been focused only on studying simple point contact CNCL which are of little practical value for industrial and engineering requirements. Thus to bridge this gap a 1-D single phase semi-analytical method to model the transient CNCL system having flat plate heat exchanger formed via the coupling of constituent square NCLs which have a square cross-section is proposed. To validate the proposed 1-D model a 3-D transient study is carried out using ANSYS Fluent 16.1 after the grid and time-step independence is established. The CNCL with flat plate heat exchanger section is studied for both the vertical and horizontal configurations, thus ensuring a thorough and in-depth study. The following conclusions can be drawn from the present work:

1. The 1-D model of the CNCL system is an ideal tool that greatly minimizes the time required to understand the physics of the system, and it is robust enough to handle all the possible heat transfer boundary conditions (constant temperature, heat flux and volume generation), transient heat inputs and heat loss to the ambient surroundings. The 1-D model accurately predicts the trend of the CNCL system with minimal computational effort, thus the current work greatly simplifies and aids in the design of the CNCL system.

2. The cross section averaged velocity of each loop and the volume-averaged temperature of each loop are excellent parameters for comparing the 1-D transient single-phase model with the 3-D CFD study for the CNCL system.

3. The transient heat transfer coefficient at the common heat exchanger section is estimated by conduction in the thermal boundary layer during the initial transience and the convective flow correlations can be used to predict the heat transfer behaviour thereafter.

4. The transient averaged fluid velocity plots are influenced by forces acting on the fluid element, which include buoyancy, viscous forces and drag forces induced by the bends. The transient average temperature plots are a strong function of the axial conduction within the fluid element and its heat storing capacity.

5. The CNCL orientation (vertical or horizontal), coupled with the heater and cooler configuration determines the system dynamics and behaviour. From the current study, we find vertical CNCL systems to exhibit less oscillations in comparison with the horizontal CNCL at low power loads but the behaviour of the system is reversed at higher loads.

6. The present study indicates that a vertical CNCL at steady state always has a counterflow heat exchange configuration at the common heat exchanger region; whereas the horizontal CNCL portrays both counter and parallel flow configurations depending on the flow initialisation and the heater and cooler position at low power loads.

7. The heat transfer coefficient at the common heat exchanger region does not play a significant role in the transient and steady state behaviour of the CNCL system when the operating fluids have low Prandtl numbers (liquid metals and fluids used for the present CFD study) .

8. The CNCL system also exhibits chaotic behaviour at higher heat flux, which needs to be avoided to ensure designed system behaviour and direct control over system dynamics.

9. The effect of expansion tank on the CNCL system behaviour is not considered in the present 1-D study.

10. The contrast in the behaviour of the Horizontal system and Vertical CNCL system can be understood in terms of the vicinity of the heat exchanger location to the vertical limbs (w.r.t gravity). It is this factor that enables the Horizontal CNCL system to have a lower average fluid velocity and a lower average temperature magnitude at steady state relative to the vertical CNCL.

11. The initial velocity imparted to the fluid reduces the time taken to attain steady state significantly for a horizontal CNCL system, relative to a zero initial velocity flow field initialisation. The initial temperature has no effect on the CNCL system apart from offsetting the transient average temperature plot curve.

12. The transient nature of the average fluid velocity and average fluid temperature of the Horizontal CNCL are different from that of an NCL system, this difference can be attributed to the variation in transient temperature profile in the cooler sections of the respective systems. The average temperature of the NCL system remains constant for the constant heat flux boundary condition at the heater and cooler sections.

13. Future work is aimed at stability analysis of a CNCL system for a better understanding of the system behaviour.